\documentclass[preprint]{aastex}
\usepackage{rotating}
\usepackage{epsfig}  
\usepackage{epstopdf} 
\usepackage{graphics}
\usepackage{listings}

\begin{document}


\title{Quantifying the Carbon Abundances in the Secondary Stars of SS Cygni,
RU Pegasi, and GK Persei}

\author{Thomas E. Harrison$^{\rm 1,2}$}

\affil{Department of Astronomy, New Mexico State University, Box 30001, MSC 
4500, Las Cruces, NM 88003-8001}

\email{tharriso@nmsu.edu}

\author{Ryan T. Hamilton}

\affil{SOFIA Science Center, NASA Ames Research Center, Moffett Field, 
Mountain View, CA 94043}

\email{rthamilton@sofia.usra.edu}

\begin{abstract}
We use a modified version of MOOG to generate large grids of synthetic spectra in 
an attempt to derive quantitative abundances for three CVs (GK Per, RU Peg, and SS 
Cyg) by comparing the models to moderate resolution (R $\sim$ 25,000) $K$-band 
spectra obtained with NIRSPEC on Keck. For each of the three systems we find solar, 
or slightly sub-solar values for [Fe/H], but significant deficits of carbon: for SS 
Cyg we find [C/Fe] = $-$0.50, for RU Peg [C/Fe] = $-$0.75, and for GK Per [C/Fe] = 
$-$1.00. We show that it is possible to use lower resolution (R $\sim$ 2,000) 
spectra to quantify carbon deficits. We examine realistic veiling scenarios and 
find that emission from H I or CO cannot reproduce the observations.
\end{abstract}


\begin{flushleft}

$^{\rm 1}$ Visiting Observer, W. M. Keck Observatory, which is operated 
as a scientific partnership among the California Institute of Technology, the 
University of California, and the National Aeronautics and Space 
Administration.\\

$^{\rm 2}$ Visiting Astronomer at the Infrared Telescope Facility, which is 
operated by the University of Hawaii under contract from NASA.\\

\end{flushleft}
\clearpage

\section{Introduction}
Cataclysmic Variables (CVs) are a diverse group of short-period binaries in 
which a late-type, Roche-lobe filling secondary star transfers matter through 
an accretion disk onto a white dwarf primary. The
standard paradigm for the formation of CVs postulates that they start out as 
wide binaries (on the order of 1 AU), of moderate orbital period ($<<$ 1 yr). 
As the more massive component, the white dwarf progenitor, evolves off
of the main sequence, the secondary star suddenly finds itself orbiting 
within the red giant photosphere. During this ``common envelope'' (CE) phase, 
much of the angular momentum of the binary is believed to be shed due to 
interactions of the secondary star with the atmosphere of the red giant. 
During this process, the binary period gets shortened, and the red giant 
envelope gets ejected. Depending on the input parameters used in standard 
CE models, between 50 and 90\% of all of these close binary stars merge (c.f., 
Politano \& Weiler 2007). After emerging from the CE phase, an epoch of angular
momentum loss via magnetic braking (Verbunt \& Zwaan 1981), lasting $\approx$
10$^{\rm -7}$ yr (Warner 1995), is then required that leads to the formation 
of a semi-detached binary, and the mass transfer phase that signals the birth 
of a CV. 

In this scenario, the vast majority of CV secondary stars should not show any 
signs of evolution, as there just has not been sufficient time for the low 
mass secondary stars (M$_{\rm 2}$ $\leq$ 1 M$_{\sun}$) in these systems to 
have begun to evolve off of the main sequence. Thus, the composition of the 
secondary stars should be relatively normal unless they have accreted matter 
with an unusual composition during the CE phase, or perhaps during classical 
novae eruptions. In an infrared spectroscopic survey of CVs, Harrison et
al. (2004, 2005) found that the CO features of the secondary stars in a 
number of CVs were much weaker than expected for their spectral types. For U 
Gem, the CO features were extremely weak, yet the water vapor feature was
normal for a spectral type of M4V. This suggested a deficit of carbon. 
Hamilton et al. (2011) have tabulated the list of CVs with weaker than 
expected CO features, and show that, when such data exists, analysis of UV 
spectra {\it also} find evidence for carbon deficiencies. One issue with 
the infrared spectroscopy surveys was that they were qualitative, and were 
derived from lower resolution (R $\sim$ 2000) spectroscopy. The line features 
in those spectra {\it might} be compromised due to veiling by a continuum 
source that arises from the accretion disk. 

We have obtained moderate, and lower resolution, $K$-band spectroscopy of 
three of the brightest CVs: SS Cyg, RU Peg, and GK Per. Previous near-infrared 
spectroscopy with SPEX (Harrison et al. 2004, 2007) revealed that the secondary 
stars in these three systems had weaker than expected CO features. In the 
following, we use a modified version of the MOOG spectral synthesis program 
to model moderate resolution $K$-band spectra to examine the atomic and 
molecular abundances in their secondary stars. As shown in Hamilton et al. 
(2011), higher resolution spectra dramatically reduce the effects of accretion 
disk veiling, allowing us to see secondary stars even in CV systems where the 
accretion luminosity totally dominates that of the secondary star. In the 
systems discussed here, however, the secondary star dominates the accretion 
disk flux in the $K$-band, and any veiling has to be quite small. We then 
compare the best fit model spectra derived from the higher resolution data 
to lower resolution data.  

In the next section we discuss the data to be modeled, in section 3
we discuss the spectral synthesis procedure, in section 4 we 
present the modeling of the data, discuss the results in section 5 and draw
our conclusions in section 6.

\section{Observations}

SS Cyg, RU Peg, and GK Per were observed using 
NIRSPEC\footnote{For more on NIRSPEC go to http://www2.keck.hawaii.edu/inst/nirspec/nirspec.html} in high resolution mode 
on Keck II on 2004 August 28. The 0.432 x 12\arcsec\ slit was used
providing a resolving power of R $\sim$ 25000, or a velocity resolution of
$\Delta v$ $\sim$ 12 km s$^{\rm -1}$. We employed the 
two-nod script, and we used five minute exposure times for the three program 
CVs. To completely cover the $K$-band using NIRSPEC requires two grating 
settings. Due to time limitations, however, we only obtained data at a single 
grating position chosen to insure coverage of the first overtone bandhead of 
$^{\rm 12}$CO$_{\rm (2,0)}$ at 2.294 $\mu$m. This grating setting also covers
the $^{\rm 12}$CO$_{\rm (5,3)}$ bandhead at 2.383 $\mu$m. In addition to the 
three program CVs, we also observed two late-type stars, $\epsilon$ Eri (K2V) 
and $\gamma$ Psc (G9III), to provide templates for our synthetic spectral 
program given that the properties of those two objects are well known. 

To correct for telluric absorption, we observed bright A0V stars located
close to the program objects so as to minimize their relative differences in 
airmass. These data were reduced using the IDL routine ``$redspec$'', specially
developed for NIRSPEC\footnote{Details about the~$redspec$ package can be 
found here: http://www2.keck.hawaii.edu/inst/nirspec/redspec/index.html}.
In the $K$-band, the spectra of A0V stars are nearly featureless, except
for the prominent H I Brackett $\gamma$ feature at 2.16 $\mu$m. The $redspec$ 
package does not attempt to correct for this feature, but can interpolate
across such lines to reduce their impact upon division into the program star 
spectrum. Note that there is a weak telluric feature located very close to 
the Brackett $\gamma$ line, and thus the H I line profiles in spectra 
produced by the division of a ``patched'' A-star spectrum are slightly 
compromised. 

Just prior to our observing run on Keck, we observed the three program CVs
using SPEX at the IRTF on 2004 August 15. SPEX was used in cross-dispersed 
mode generating spectra spanning the range from 0.8 to 2.45 $\mu$m. In the 
$K$-band, the dispersion is 5.4 \AA/pix. The raw data were reduced using
SPEXTOOL (Cushing et al. 2004). A log of all of our observations is 
presented in Table 1, and includes the orbital phase at the mid-point of
the observational sequence for each of the three CVs. The AAVSO light curve 
data base\footnote{https://www.aavso.org/data/lcg} shows that SS Cyg was in 
quiescence at the time of both the SPEX 
and NIRSPEC observations, having reached visual maximum on 17 July. RU Peg 
was in outburst on 23 July, but was in quiescence for both observing runs. At 
the time of the infrared observations, GK Per had $V$ = 13.1, its normal 
minimum light value.

\section{Spectral Synthesis and Stellar Atmosphere Modeling}

To calculate a synthetic spectrum, three basic building blocks are required: A
stellar atmosphere model, a list of atomic and molecular lines, and
a tool to calculate the corresponding flux at the stellar surface as a
function of wavelength. A full description of the modified
version of MOOG and the grid generation and spectral matching Python code 
(called ``{\it kmoog}''\footnote{http://astronomy.nmsu.edu/tharriso/kmoog}) used below is described in Hamilton (2013), here we 
highlight the most important aspects.

\subsection{Model Atmospheres}

There are several groups that provide precomputed atmosphere models, each with
their own slightly different assumptions, simplifications, and basic input
physics.  Most of the models in general circulation, however, have the same
underlying assumptions: The atmosphere is either modeled as a series of 
plane-parallel slabs (1D), or as a series of concentric shells (2D). The 1D 
and 2D atmosphere models are calculated assuming local thermodynamic 
equilibrium (LTE) conditions, in which one temperature can describe the 
particle velocity distribution as well as the population of atomic levels and 
ionization states around a given point in the stellar atmosphere. It is also 
assumed that the atmosphere is time invariant, with the effects of dynamical 
processes such as mixing and convection parameterized. Bonifacio et al. (2012) 
provide an overview of the more detailed assumptions and differences between 
the available atmosphere computation methods and programs. Recently, full 3D 
(radiation-hydrodynamic) stellar atmosphere models have been calculated
across parameter grids (e.g., Ludwig et al. 2009, Hauschildt and Baron 2010, 
Magic et al. 2013a, 2013b). For this program, however, we have only used 
the 1D and 2D models. 

The most commonly used atmosphere model codes are PHOENIX\footnote{http://www.hs.uni-hamburg.de/EN/For/ThA/phoenix/index.html} (Hauschildt et al. (1999), 
Allard et al. 2011), MARCS\footnote{http://marcs.astro.uu.se/} (Gustafsson
et al. 1975, 2008), and ATLAS\footnote{http://kurucz.harvard.edu/programs.html} 
(Sbordone et al. 2004). PHOENIX is perhaps the most computationally 
sophisticated, sampling 42 million atomic and 550 million molecular lines 
when calculating opacity. MARCS model atmospheres are consistently updated 
through a web interface providing both 1D and 2D models. Ultimately for this 
project, we chose to use MARCS atmospheres. An important 
consideration for this choice was the existence of the interpolation code to 
create subgrid models, making it much simpler to construct a large grid of 
models quickly. Very recently, M\'{e}sz\'{a}ros et al. (2012) presented a new 
grid of atmospheres based off of ATLAS and MARCS codes with an updated H$_2$O 
line list as well as finer grid spacings in temperature, gravity, metallicity, 
and C and $\alpha$ abundances. Given their recent availability, however, all 
analyses and spectra computed here use solely the original MARCS grid.

\subsection{Constructing an Input Line List}

The next crucial step to constructing a synthetic spectrum is to assemble a
list of atomic and molecular transitions for which you wish to calculate an
opacity at the stellar surface. For our project we used the ATLAS line list 
with the addition of the CO line list by Goorvitch (1994). Each line list was 
converted into a common format before combining, and, in the case of the CO 
line list, only lines with $log_{10}(g_\mathrm{f}) \geq -16$ and 2.28 
$\leq \lambda \leq$ 2.405 $\mu$m were kept. The final list contains a total of 
11,681 lines. For the purposes of this work, only computations involving 
diatomic molecules were considered. This eliminates lines of the important 
molecular species for the cooler M dwarfs commonly found in CVs, such as 
H$_2$O. As described below, synthetic spectra for the Sun and Arcturus were 
generated and compared to observations to provide the initial validation of 
this line list.

\subsection{Spectral Synthesis Tool}

There are a number of codes that can readily make use of MARCS and Kurucz
atmospheres. MOOG\footnote{http://www.as.utexas.edu/$\sim$chris/moog.html} 
(Sneden 1973) has one of the longest track records, and given its ubiquity,
we chose it to compute the synthetic spectra for this work.  
When creating a spectrum, MOOG is internally limited to considering only
2500 lines in the line list. Given the size of our line list, we modified
MOOG (see Hamilton 2013) to allow for lists of 250,000 lines. Two additional 
changes were made to the MOOG source code. First, a version was created to 
produce absolutely no output to the screen. This prevents unnecessary lag and 
slowdowns, as well as greatly enhances usability when running a large number 
of models simultaneously. Second, and most importantly, a new binary output 
file type of MOOG was created. Significant savings in storage size were 
realized by creating a true FITS output type, reducing the file size of a
successfully created model by a factor of five or more. The FITS files 
created also have the benefit of being fully compliant and compatible with
the standard CFITSIO\footnote{http://heasarc.gsfc.nasa.gov/fitsio/}
libraries, unlike the FITS-like output already in the code.  The FITS output
created also has the benefit of storing both a full resolution 
spectrum as well as one convolved to lower resolutions, useful for quickly
comparing against observations from different telescopes and instruments.

\subsection{Validation of the Synthetic $K$-band Spectra}

To validate our line list, we constructed synthetic spectra for the Sun
(using the solar MARCS model) and Arcturus (using the parameters listed in 
Table 2, but with log$g$ = 3.0) to compare them to high resolution $K$-band 
spectra published by Wallace et al. (1996), and Wallace \& Hinkle (1996). We 
concentrated on the reproduction of only the strongest spectral features 
observed in these two objects. Note that the Arcturus comparison was only
used to insure that we were not missing lines that become important at
lower temperatures. We then altered the oscillator strengths of the 
most highly discrepant lines to insure that they would both fit the 
observations better, and not bias the automated fitting process. For example, 
there were a
large number of Sc I transitions that had oscillator strengths that were
much too large. Unfortunately, oscillator strengths for many of the
transitions in the $K$-band do not appear in the NIST Atomic Spectra 
Database\footnote{http://physics.nist.gov/PhysRefData/ASD/lines\_form.html}, 
and thus we used trial and error adjustment of the tabulated oscillator 
strength in the line list to better reproduce the observed spectral feature
(for a primer on the identification of the strongest spectral features found 
in the $K$-band for K-type stars, see Fig. 5 in Harrison et al.  2004).

To test our ability to reproduce observations, we have constructed
synthetic spectra to compare to the NIRSPEC observations of $\epsilon$ Eri and
$\gamma$ Psc, and to a sample lower resolution spectra of K dwarfs contained 
in the IRTF Spectral Library\footnote{http://irtfweb.ifa.hawaii.edu/~spex/IRTF\_Spectral\_Library/} (Cushing et al. 2009). We summarize the spectroscopically 
derived parameters for these objects in Table 2, averaging the data compiled in 
the PASTEL catalog (Soubiran et al. 2010) for each object, constrained to 
measures made after 1970. The final column of Table 2 denotes how many
unique measurements went into the determination of each of the average properties 
for each star (note that HD45977 only has a single measurement of its 
characteristics by Sousa et al. 2011). As we will see, this sample more than 
spans the range in metallicity observed for our program CVs. Note that the solar 
abundance pattern derived by Grevesse et al. (2007) was used for all of the 
synthetic spectra, and only the abundance of carbon was altered for the  
investigation of the CVs.

In preparation for comparison to the models, the spectra were continuum 
normalized using a third order polynomial fit in IRAF. For the high resolution
spectra, the process was quite trivial. For the full $K$-band spectra, however,
we used an identical set of wavelength bins for identifying the continuum
to be fit. The continua were then fitted over the range 2.08 to 2.29 $\mu$m, 
with bins that avoided all of the strongest spectral features.

\subsubsection{$\epsilon$ Eri and $\gamma$ Psc}

\begin{figure}[htb]
\centerline{{\includegraphics[width=15cm]{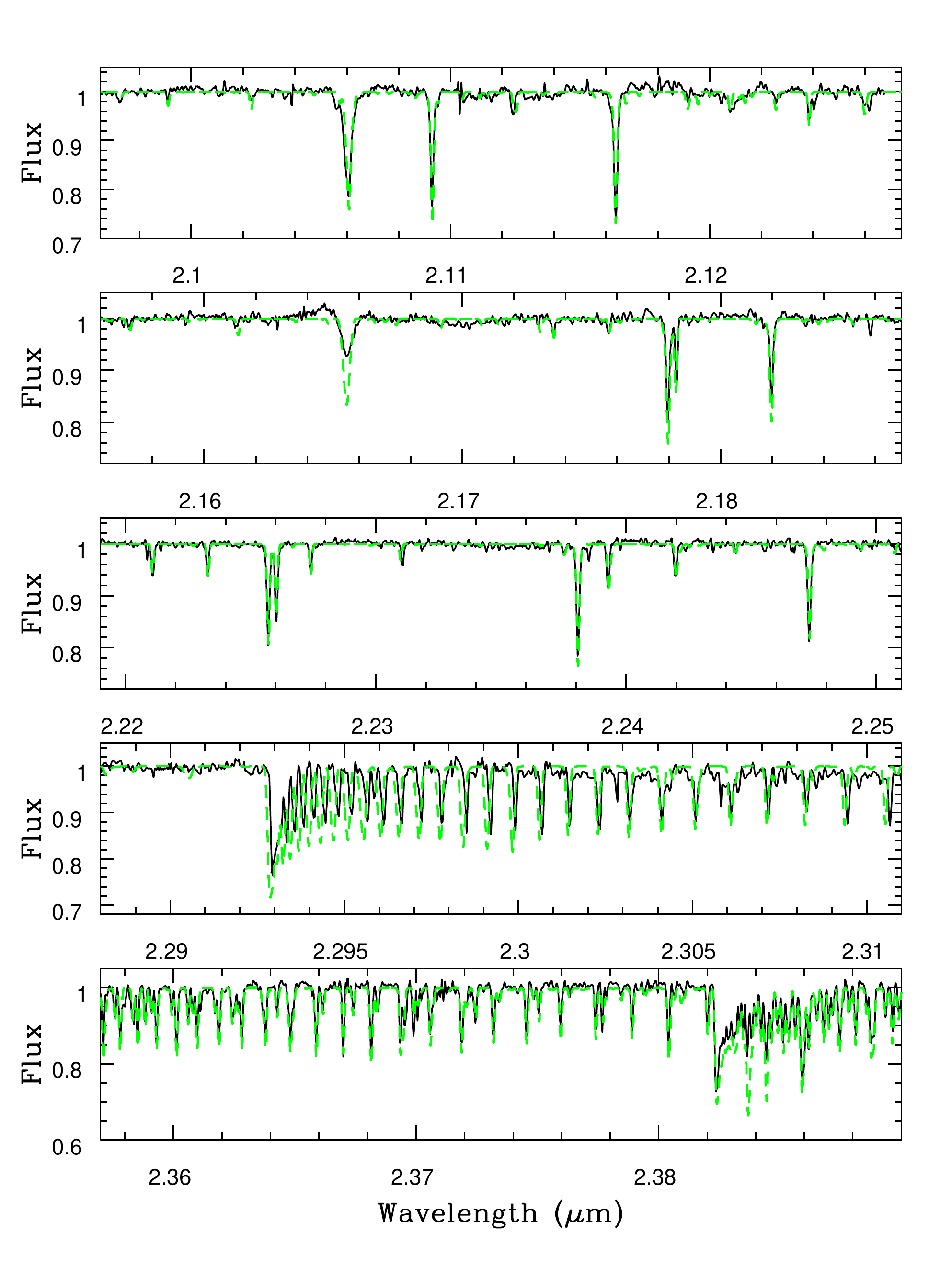}}}
\caption{The NIRSPEC spectrum of $\epsilon$ Eri (black solid line). The 
synthetic model spectrum (green, dashed line) generated for this object is 
overplotted. For the purposes of this paper, the bottom panel will
be defined as the second order, while the top panel will be labeled the sixth 
order. The synthetic spectrum has been rotationally broaded by 4 km 
s$^{\rm -1}$ as measured by Martinez-Arnaiz et al. (2010). The S/N ratio
ranged from 214 in the sixth order, to 95 in the second order.}
\label{eps}
\end{figure}

The NIRSPEC spectrum of $\epsilon$ Eri is presented in Fig. \ref{eps},
where data for five of the seven spectral orders are shown (the first order, 
spanning $\lambda$2.43 to 2.47 $\mu$m, was too far red to obtain appreciable 
signal in our CV spectra, while the seventh order spectrum is dominated by the 
strong telluric feature at 2.05 $\mu$m). As shown in Table 2, $\epsilon$ Eri 
has a temperature near 5100 K, and a near-solar metallicity of [Fe/H] = $-$0.1.
In Fig. \ref{eps}, we overplot a synthetic spectrum with T = 5100 K, and
[Fe/H] = $-$0.125 (interpolated between the model atmospheres with [Fe/H] = 
$-$0.25, and [Fe/H] = 0.0). The model fit to the strongest lines is excellent.
Note that we have inserted an H I Br$\gamma$ line into our line list. We do this
to primarily show where this line is located. The profiles of this feature
as found in the infrared spectra for each of the objects are unreliable due to 
the reduction technique. In addition, the profile of such a strong line cannot 
be easily reproduced using MOOG.

\begin{figure}[htb]
\centerline{{\includegraphics[width=15cm]{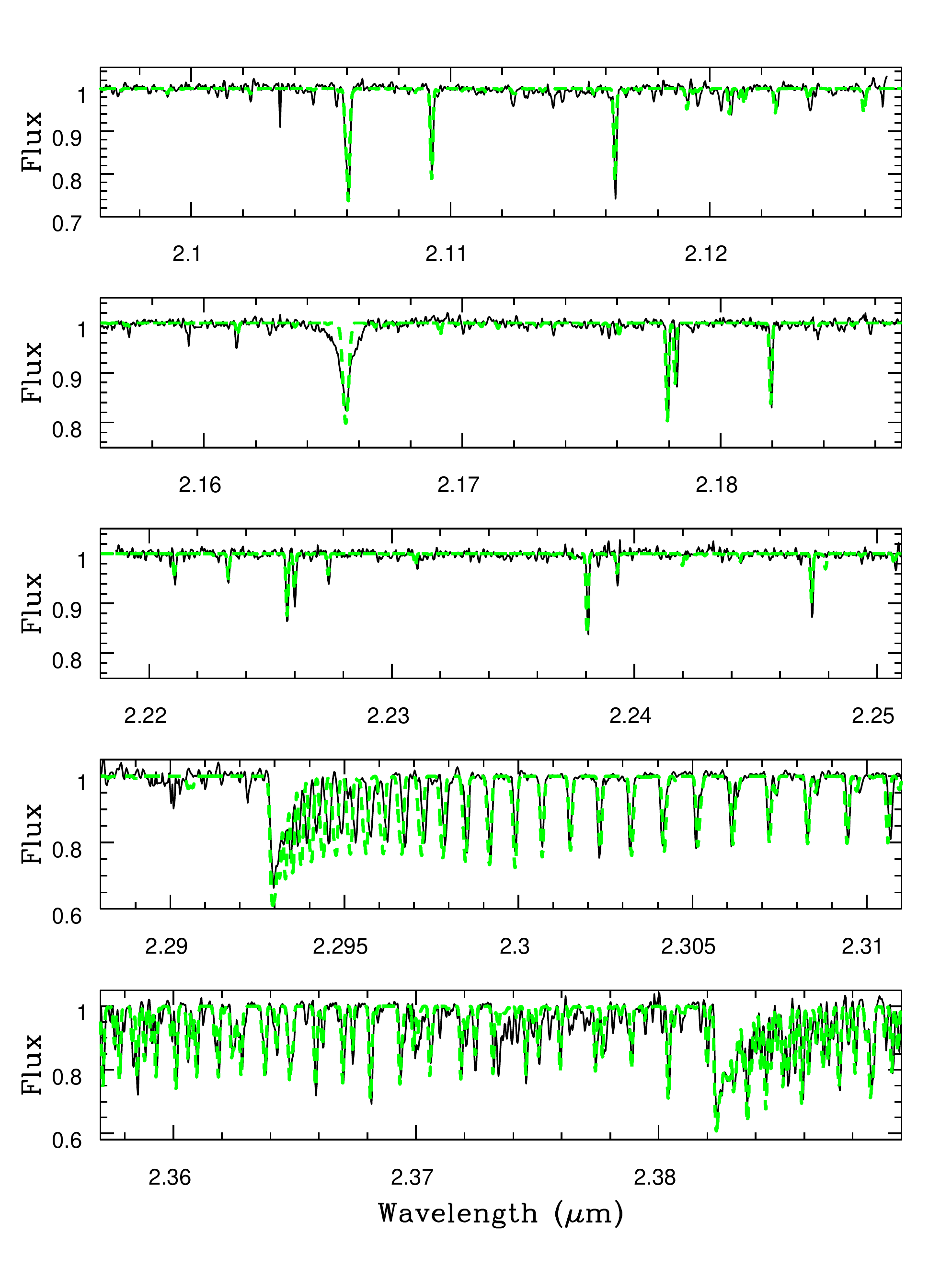}}}
\caption{The NIRSPEC spectrum of $\gamma$ Psc (black) compared to
a synthetic spectrum (green). The S/N ratio of these data
ranged from 175 in the sixth order, to 75 in the second order.}
\label{gamma}
\end{figure}

$\gamma$ Psc allows us to test our ability to reproduce a much lower gravity 
object with a very low metallicity. We used a MARCS spherical, alpha-enhanced
model atmosphere with log$g$ = 2.5, and [Fe/H] = $-$0.5 to generate a synthetic
spectrum for $\gamma$ Psc. As noted in Alves-Brito et al. (2010), $\gamma$ Psc
has a modest alpha-enhancement of $\sim$ $+$0.2, and thus many of the metal lines are 
stronger in $\gamma$ Psc than they would be in a model that has a solar abundance 
pattern. Since $kmoog$ does not yet have the capabilities to adjust the abundances 
of any other species besides carbon, we ran the MOOG binary used by $kmoog$ in a 
stand-alone mode, increasing the abundances of O, Al and Mg as detailed in 
Alves-Brito et al. We overplot this model on the observed data in Fig. \ref{gamma}. 
The resulting synthetic spectrum is a reasonable match to the data.

\subsubsection{Comparison to the K Dwarfs in the IRTF Spectral Library}

\begin{figure}[htb]
\centerline{{\includegraphics[width=15cm]{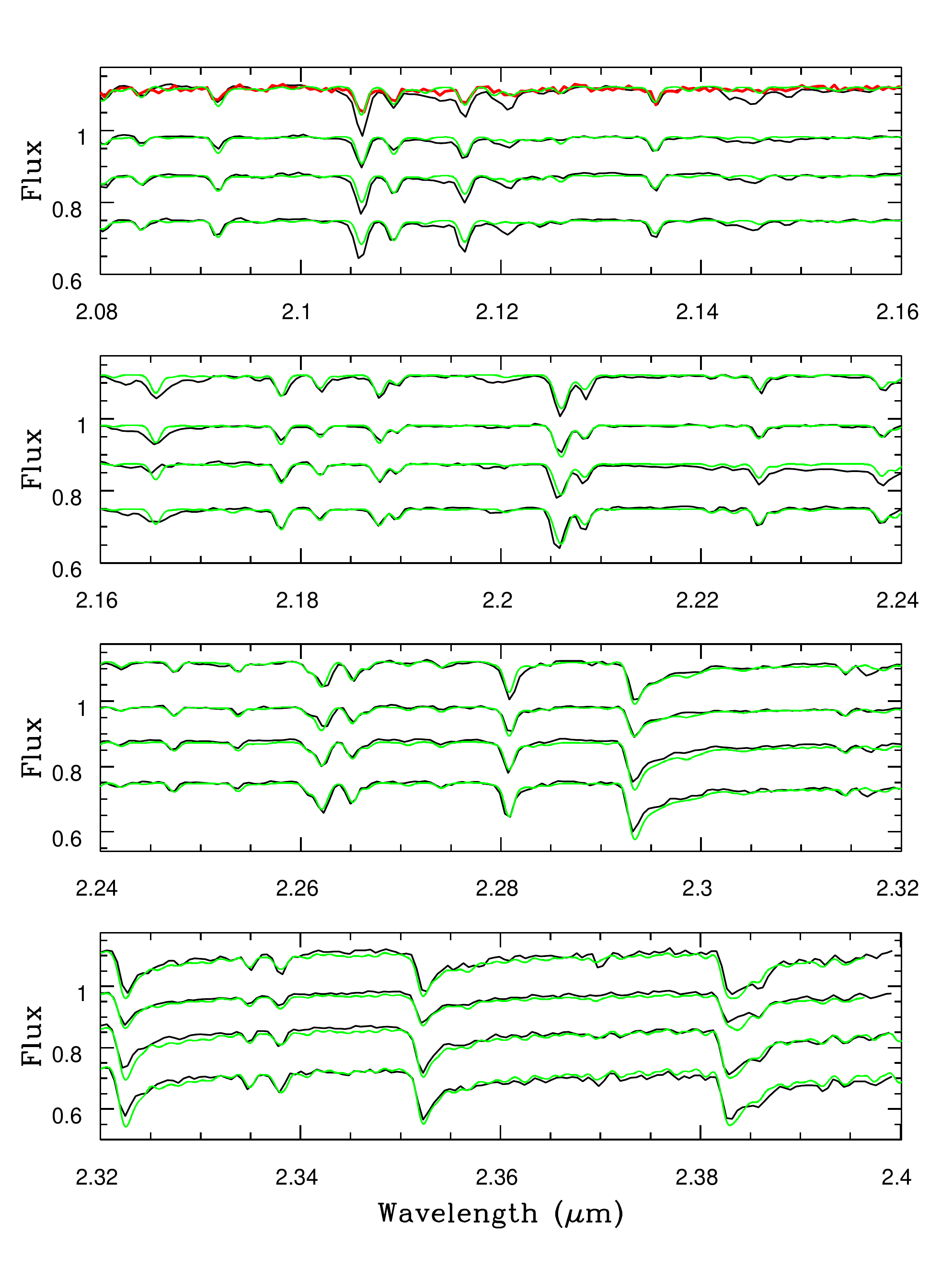}}}
\caption{\small The SPEX spectra for K dwarfs spanning the range from K0V to
K3V (black), along with synthetic spectra (green) generated for each of
the objects using the parameters listed in Table 2. As described in the text,
the template star spectra in the range 2.10 to 2.15 $\mu$m are not especially
well matched by the models. The source of this discrepancy is hard to identify.
In the top panel we also plot in red the spectrum of a G9V star (HD161198) 
that we observed with SPEX during the 2004 August observing run, and the model 
spectrum generated for the K0V template fits the spectrum of HD161198 quite
well.}
\label{k0tok3}
\end{figure}

\begin{figure}[htb]
\centerline{{\includegraphics[width=15cm]{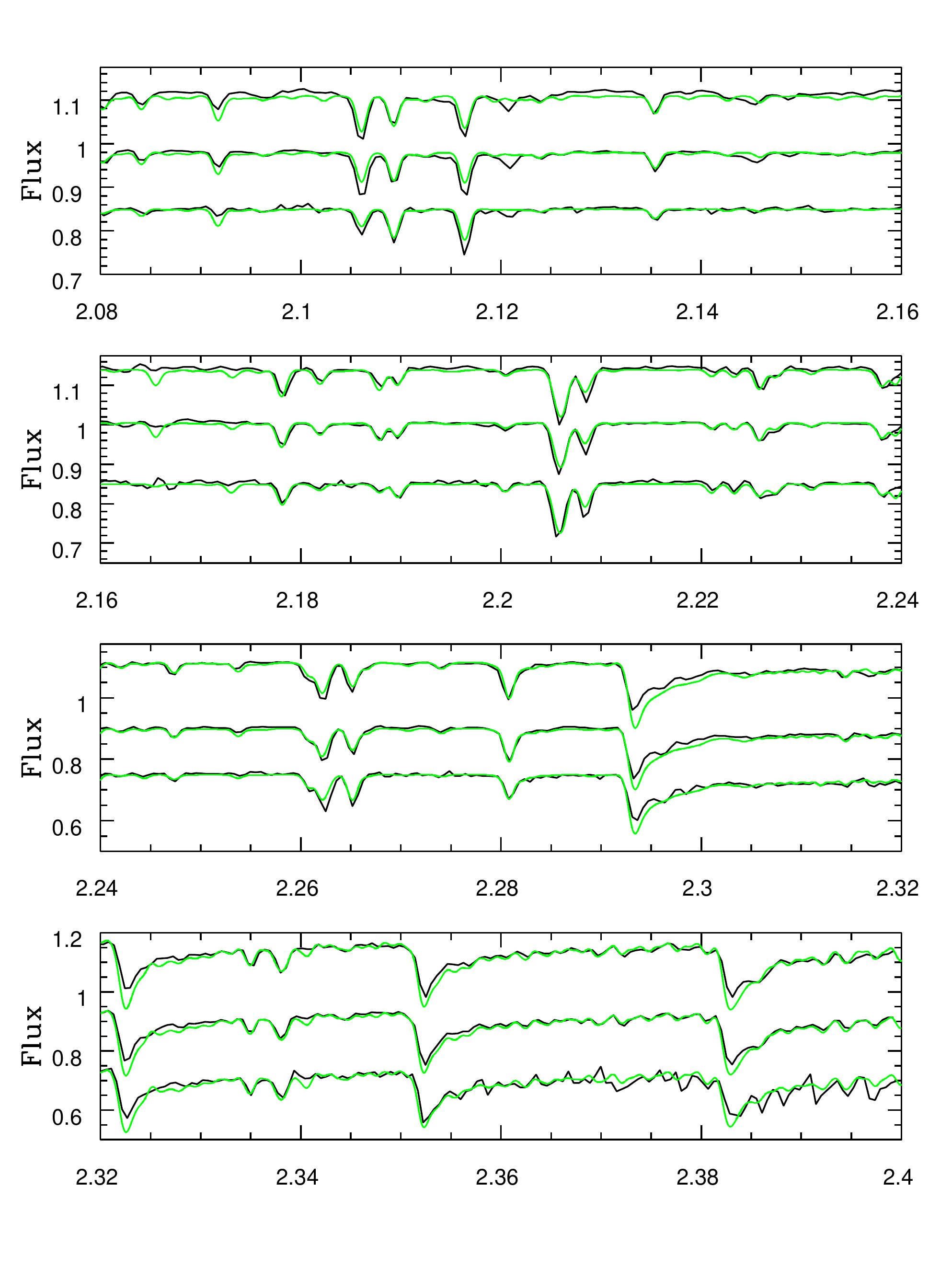}}}
\caption{The SPEX spectra for K dwarfs spanning the spectral types from K4V to
K7V (black), along with synthetic spectra (green) generated for each of
the objects using the parameters listed in Table 2. For these stars,
the model spectra are somewhat better at matching the observed spectra in
the 2.10 to 2.15 $\mu$m region. }
\label{k4tok7}
\end{figure}

Given that the majority of CVs are too faint for moderate resolution infrared
spectroscopy, most of the existing $K$-band data have been obtained at a
lower resolution. To examine how well we can model such data sets we
have downloaded SPEX data for dwarf stars spanning the spectral types K0V to 
K7V. The spectral type range K0V to K3V is shown in Fig. \ref{k0tok3}, while
the range K4V to K7V is shown in Fig. \ref{k4tok7}. We generated synthetic
spectra for each of these stars using the parameters listed in Table 2.
The published metallicities produced excellent fits over most of the spectral
range for all of the stars except HD145675 (K0V). We found that the best 
fitting model for HD145675 over the entire $K$-band required a lower
metallicity, [Fe/H] = $+$0.25, than tabulated for this object (though within
the observed error bars).

Close inspection of Fig. \ref{k0tok3} shows that the synthetic spectra do
not do a very good job of matching the strongest spectral features in the
2.10 to 2.15 $\mu$m bandpass. The origin of this discrepancy is not obvious.
The fit of the synthetic spectra over the rest of the $K$-band suggests
the tabulated (or derived) metallicities are correct. This might be
a problem with continuum subtraction. To investigate this, we reduced
and plot the SPEX data for the G9V star HD161198 observed during the same run
as our program CVs [note that HD161198 is a spectroscopic binary with
a cool secondary star (Duquennoy et al. 1996), and it is not a useful spectral 
template]. The model for HD145675 does a good job at fitting
the spectrum of HD161198 in this range, indicating that continuum subtraction
is not the issue. This suggests that the problem could be due to either the
data reduction process, or some issue with the telluric standard and/or
object spectra. The maximum amount of observed signal occurs in this
bandpass, and the K0V, K1V, and K3V have $K$ $\leq$ 4.7 (HD3765 has $K$ =
5.16, while HD145675 has $K$ = 5.55), so perhaps it is a linearity issue.
The discrepancy between the models and observed spectra in the 2.10 to 2.15 
$\mu$m bandpass is not as severe for the K4V to K7V templates (even though
the K7V template is 61 Cyg B, with $K$ = 2.32).

Clearly, however, our synthetic spectra do an excellent job at
matching objects with known temperatures and metallicities. Especially
important for the current program is the match of the model spectra
to the observed CO features over a wide range of temperature and 
metallicity.

\subsection{Creation of a Large Parameter Grid}

In the preceding we have modeled the spectra of objects with known parameters.
For modeling the CVs, where the temperatures and metallicities are poorly
known, requires the generation of a large number of models, and a robust method
to match the synthetic spectra to the observations. With the Python wrapper
we have constructed to the modified version of the MOOG program described 
above, we are able to generate a large grid of models covering a user-defined 
range of the following parameters: T$_{\rm eff}$, log $g$, [Fe/H], and [C/Fe]. 
Microturbulence was fixed at 2 km s$^{-1}$, and macroturbulence was not 
included since it is a small effect at the resolution of our observational 
data. In {\it kmoog}, if the synthetic spectra are designated to be stored 
as FITS files, each model actually contains two synthetic spectra: one 
unsmoothed, full resolution model, and one model convolved with a rotationally 
broadened spectrum\footnote{Using equation 17.12 in Gray (1992).}
appropriate for the target at the observed spectral resolution (which
must be specified before the grid creation). A full grid covering the
entire parameter space (3800 $\leq$ T$_{\rm eff}$ $\leq$ 5200 K, 3.0
$\leq$ log $g$ $\leq$ 5.0, $-1$ $\leq$ [Fe/H] $\leq$ $+$1, and $-1$ $\leq$ 
[C/Fe] $\leq$ $+$1) requires about one day to create on a 32 core machine and
occupies $\sim$  86 GByte in FITS output format. Obviously, smaller grids
can be generated much more rapidly if the input parameter space can be 
narrowed. 

\subsection{Searching the Grid: Brute Force and Genetic Algorithms}

With such large grids, it is easy to be overwhelmed by the quantity of 
data. Given a suitable subset of parameters, a direct grid
search is possible; gather a list of valid models, and step through them one
at a time comparing the model to observations.  Given the sometimes low S/N
of the observational data, however, a $\chi^2$ statistic is not calculated
across the entire observational range, but restricted to the areas of
greatest impact. For example, including Na I, Ca I, Mg I, Al I, Fe I lines, 
and the CO bands. This restricted region has the advantage of avoiding areas 
prone to increased levels of noise, and generally improves the quality of the 
fits to the observations.

Another method of stepping through the parameter space involves the use of
an evolutionary algorithm to find the best fit to the data. Evolutionary 
algorithms are based off the ideas of biological evolution, the simple 
premise that in some given population only the ``fittest individuals'' 
(combination of all parameters) will ``breed'' and influence the next 
generation of that population.  If we imagine representing our parameters 
as 1D binary strings, then the behavior and the manipulation of
the binary strings over the course of the fit is not unlike that of DNA during
reproduction.  Sets of parameters can be mutated, introducing diversity into 
the population which helps to avoid getting stuck in a local maximum or 
minimum in the ``$\chi ^{\rm 2}$ landscape''.  Mutations take the form of bit 
flips or swaps, analogous to transcription errors in DNA replication. We used 
a genetic algorithm framework called 
$PyEvolve$\footnote{http://pyevolve.sourceforge.net/}. 

The algorithm proceeds as follows:

\begin{enumerate}
\item Make an initial first guess at the parameters based on spectral shape
\item Choose an appropriate maximum number of generations to evolve through as
well as the initial population size
\item ``Breed'' and ``mutate'' the ``fittest'' members in the population to
create a new generation 
\item Repeat \#3 until the population converges to the desired level
\end{enumerate}

In this manner, all parameters are varied continually as the population
converges towards some minimum $\chi^2$, at least a factor of four faster
than the plain grid searching technique. Using both grid search and genetic 
algorithm techniques, the grid created with MOOG can be readily compared with 
the appropriate subset of CV spectra to attempt to determine as many stellar 
parameters as possible. Both methods are available in {\it kmoog}, though
we find the brute force is sufficiently rapid and as reliable as the genetic
algorithm technique. Heat maps showing the best fitting models for any
two parameters are then generated to allow the determination of the best
fitting model(s).

As we will find below, determining the exact value T$_{\rm eff}$ or [Fe/H]
for an unconstrained object from a $K$-band spectrum is difficult. Many of the 
relevant oscillator strengths for transitions in this bandpass are unknown,
and we have had to estimate a number of them by comparison of models to
observations. In addition, our line list
is incomplete, and thus weaker spectra features can contribute in a negative
fashion to the assessment of the fit of a model to the data. The issues
of telluric correction and continuum subtraction, as noted above, can 
introduce additional uncertainties into the goal of finding the best fitting 
solution. Thus, the $\chi^2$ analysis just described must be done carefully,
focussing on the strongest spectral features, insuring that there is not a 
spectroscopic anomaly in that feature which can be attributed to random 
noise variation, or to the telluric correction process. It is extremely 
difficult to produce a fully automated best fit regimen that can blindly solve 
for the parameters of a previously unconstrained object in the $K$-band, even 
when focused on specific spectral lines. Currently, we manually
examine the $\chi^2$ analysis for each line/region to insure that
it is reliable, and that no unexpected events have occurred that affect 
the line/region under analysis. One such anomaly will be encountered shortly
when we discuss the Al I line (at 2.116 $\mu$m) in SS Cyg. It has an 
unusual strength that defies obvious explanation. $\chi^{\rm 2}$ analysis 
of this line results in a super-solar abundance. We continue to work on the 
matching aspects of {\it kmoog}, but in its current state, a more interactive 
(``hands-on'') approach is required to insure the most robust results.

\section{Results}

Given that we have some information on the spectral types of the secondary
stars in our CVs, we decided to use smaller grid than can be generated with 
the default settings of {\it kmoog}. For example, tests in Hamilton (2013) show 
that even the moderate resolution $K$-band spectra are inadequate for 
assessing gravity. Given the large rotational broadening of our targets, it 
will be impossible to constrain the surface gravity for the program CVs. Thus 
we fixed log $g$ = 4.5 for modeling SS Cyg and RU Peg, and log $g$ = 4.0 for 
GK Per (see below), to reduce parameter space, and increase the efficiency of 
the grid searching process. We postpone the discussion of error bar
estimation to section 4.4.

\subsection{SS Cyg}

SS Cyg is the prototype long period (P$_{\rm orb}$ = 6.603 hr) dwarf nova
with eruptions that recur on a monthly basis. Harrison et al. (2004)
estimate the spectral type of the secondary as K4/K5, consistent with
other published values. They noted, however, that the CO features were
very weak for this spectral classification. Bitner et al. (2007) have used 
moderate resolution visual band spectroscopy to examine the masses of both 
components, and found that the secondary star is 10 to 50\% larger than
a star of its mass and temperature should be. They suggest that the secondary
star in SS Cyg is evolved, and may have a core that is depleted in hydrogen.
They also measured a rotational velocity of  $v_{\rm rot}$sin$i$ = 89 km 
s$^{\rm -1}$ for the secondary star.

\begin{figure}[htb]
\centerline{{\includegraphics[width=15cm]{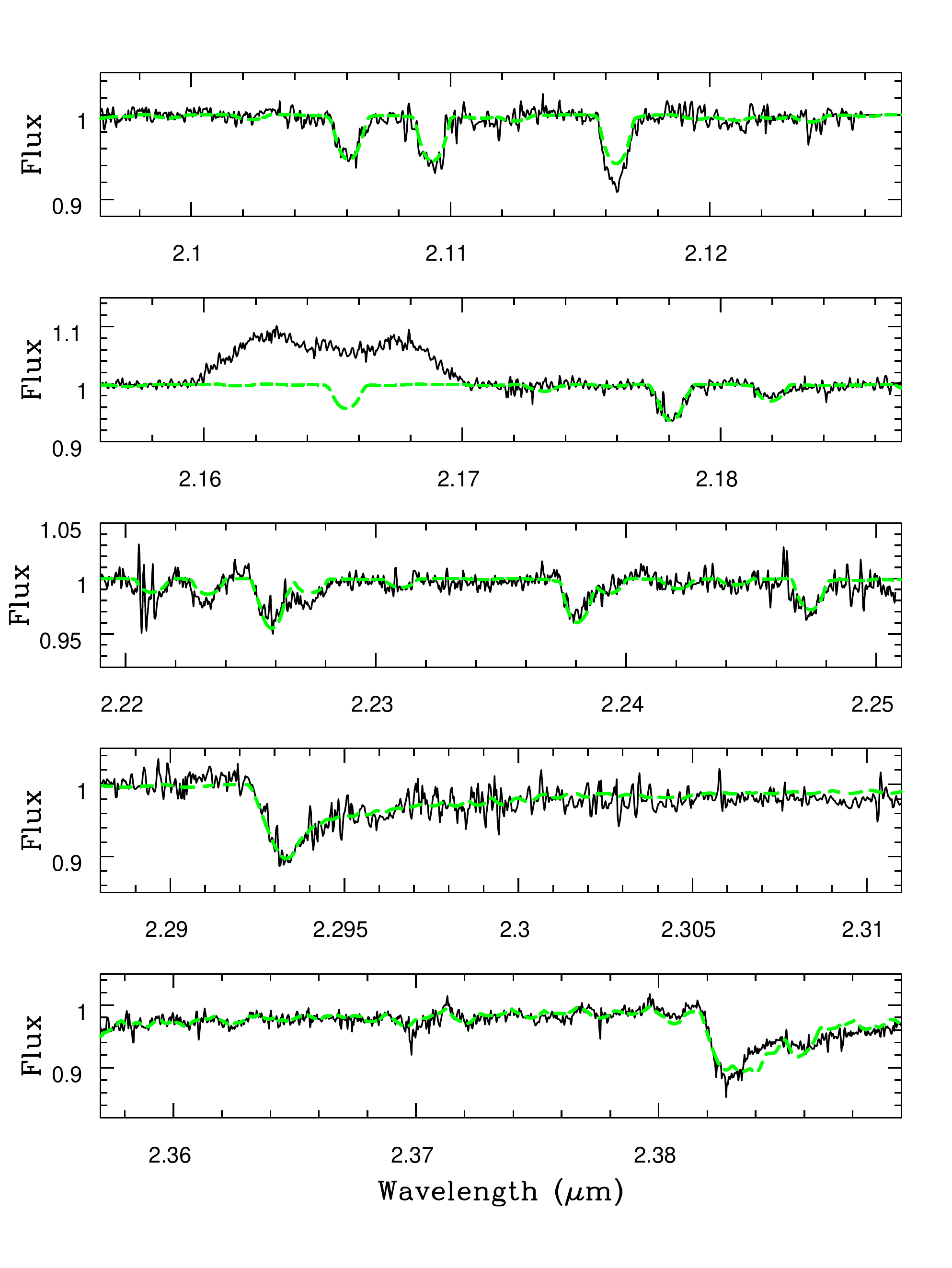}}}
\caption{The NIRSPEC spectrum of SS Cyg (black) with the final best
fitting synthetic spectrum overplotted (green). The template spectra have
been rotationally broadened by 89 km s$^{\rm -1}$. The broad emission
feature at 2.165 $\mu$m is H I Br$\gamma$. The S/N ratio in these data
ranged from 171 in the sixth order, to 67 in the second order.}
\label{sscyg}
\end{figure}

The NIRSPEC spectrum of SS Cyg is presented in Fig. \ref{sscyg}. The 
large rotational broadening of the spectral features is apparent. Our
observations spanned $\Delta \phi$ = 0.05. At the observed phase (using
the K$_{\rm 2}$ = 162.5 km s$^{\rm -1}$ value derived by Bitner et al.),
the total change in radial velocity over our set of observations would be 
$\Delta v$ = 11 km s$^{\rm -1}$, resulting in minimal ``orbital smearing.''
Concentrating on the sixth order bandpass (2.095 to 2.128 $\mu$m), we 
convolved solar metallicity models having T = 4700 K with rotational velocities 
ranging from 70 to 110 km s$^{\rm -1}$. The lowest $\chi ^{\rm 2}$ values were 
for $v_{\rm rot}$sin$i$ = 90 $\pm$ 10 km s$^{\rm -1}$. For the modeling below, 
we adopt the Bitner et al. value of $v_{\rm rot}$sin$i$ = 89 km s$^{\rm -1}$ 
for SS Cyg.

\begin{figure}[htb]
\centerline{{\includegraphics[width=15cm]{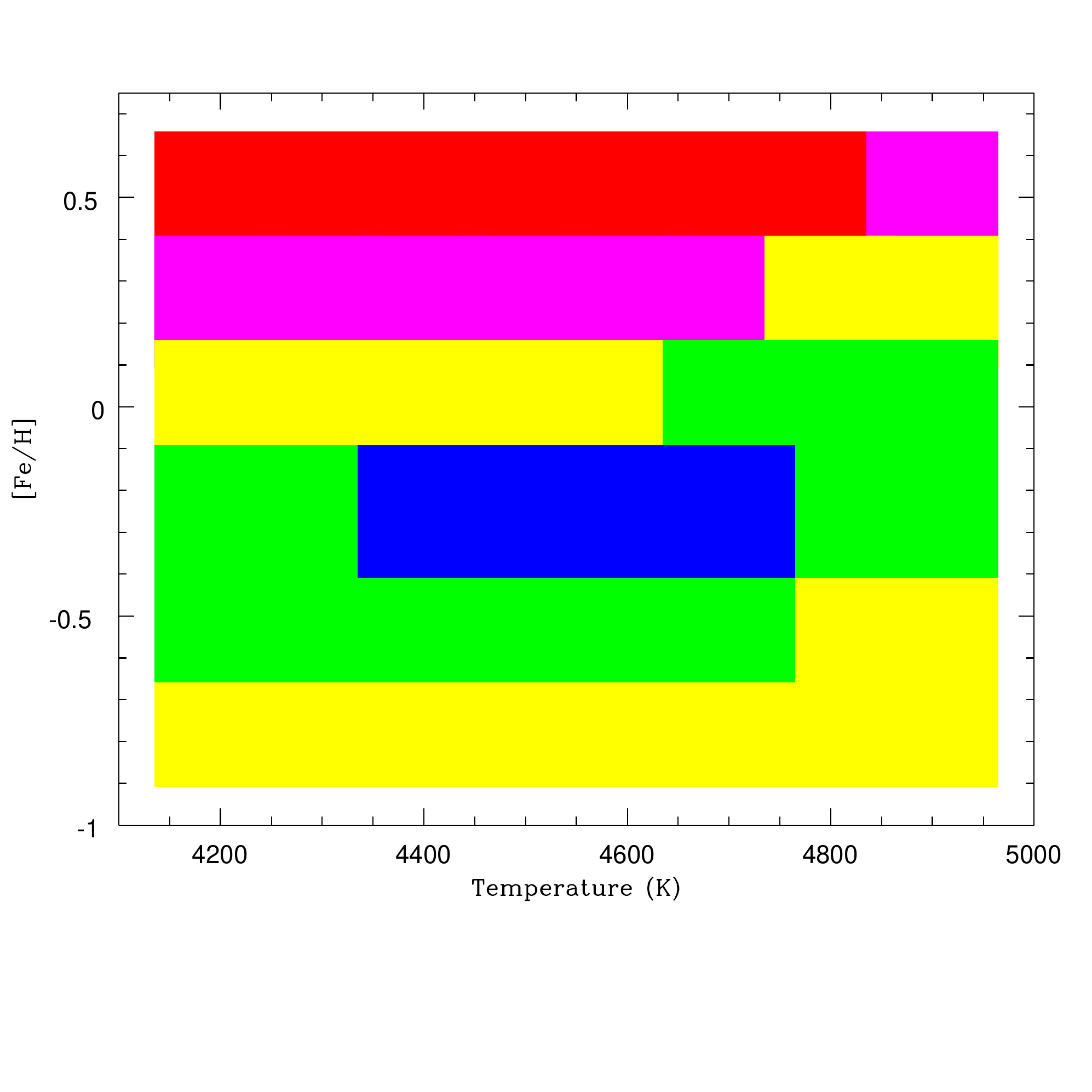}}}
\caption{The $\chi^{2}$ heat map for the order 6 spectral region
for the NIRSPEC spectrum of SS Cyg. In this, and all subsequent
heat maps, blue denotes $\chi^{2}_{\rm red}$ $<$ 0.7, green is
0.7 $\leq$ $\chi^{2}_{\rm red}$ $<$ 0.9, yellow is 0.9 $\leq$ $\chi^{2}_{\rm red}$ $<$ 1.5, magenta is 1.5 $\leq$ $\chi^{2}_{\rm red}$ $<$ 2.5,
and red is $\chi^{2}_{\rm red}$ $\geq$ 2.5. }
\label{sshm02}
\end{figure}

\begin{figure}[htb]
\centerline{{\includegraphics[width=15cm]{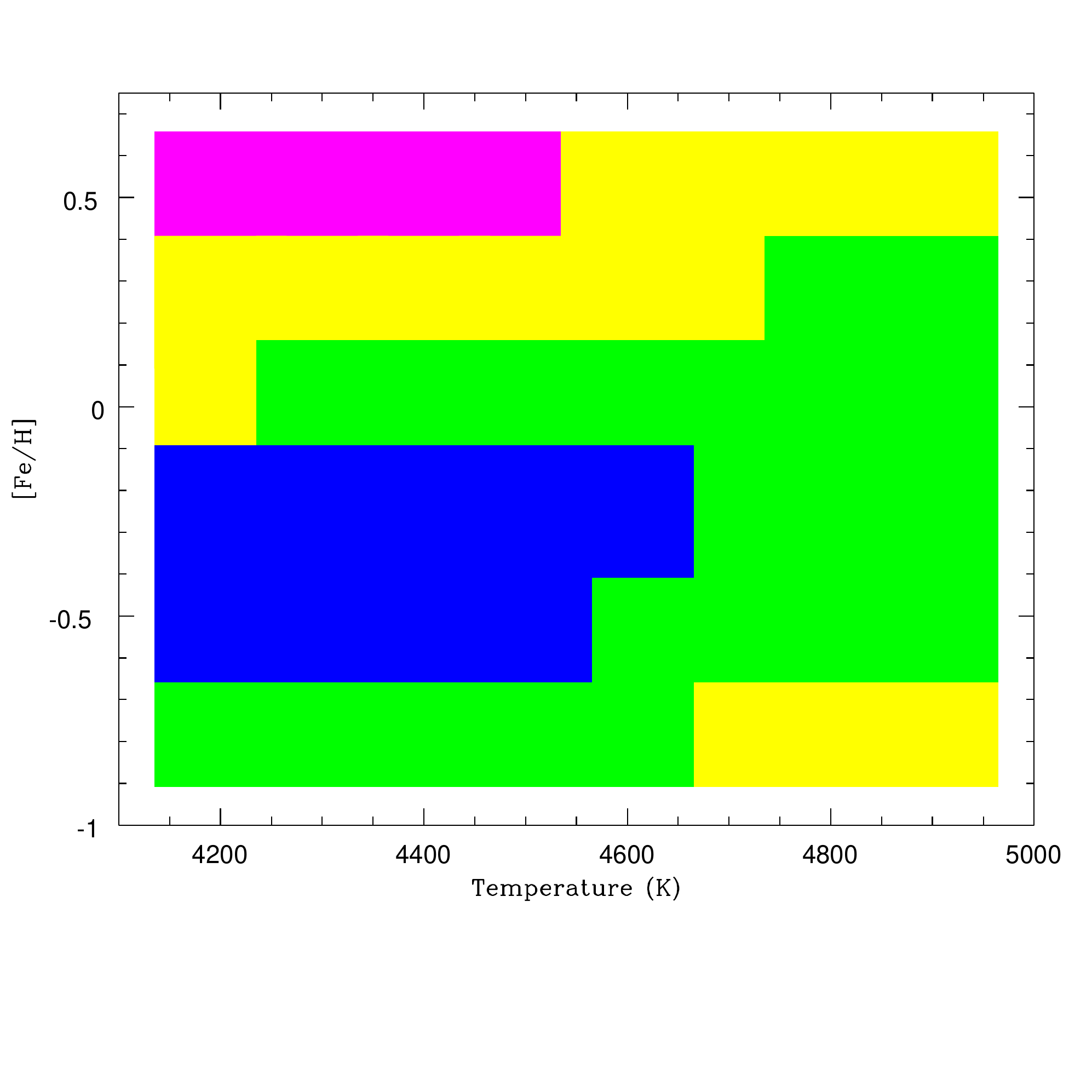}}}
\caption{The $\chi^{2}$ heat map for the order 4 spectral region 
for the NIRSPEC spectrum of  SS Cyg. }
\label{sshm04}
\end{figure}

\begin{figure}[htb]
\centerline{{\includegraphics[width=15cm]{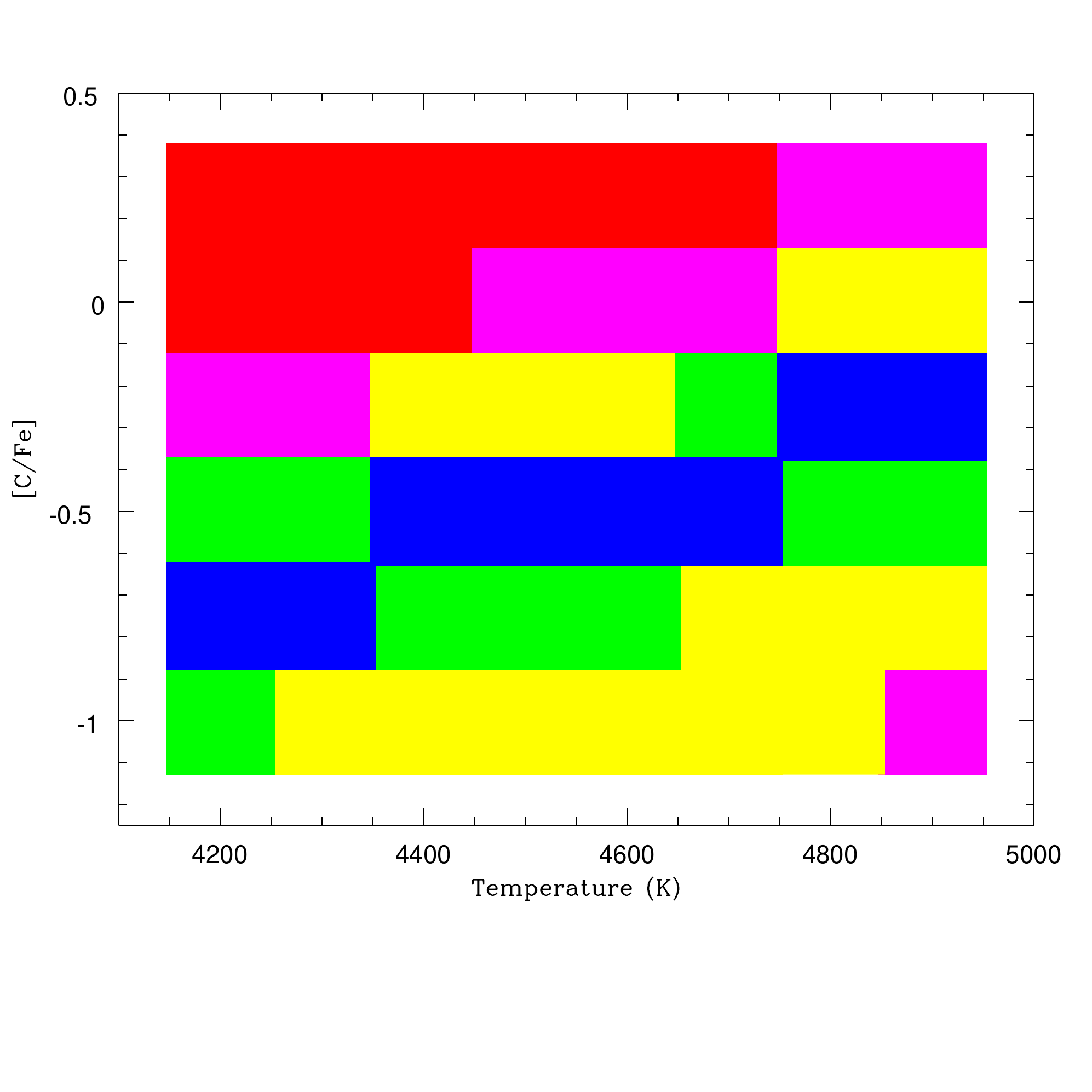}}}
\caption{The $\chi^{2}$ heat map for the order 3 spectral region
for the NIRSPEC spectrum of SS Cyg. For this analysis, a grid of
models with the metallicity fixed to [Fe/H] = $-$0.25 was constructed,
while the carbon abundance was altered over the range $-$1.0 $\leq$ 
[C/Fe] $\leq$ $+$0.25. }
\label{sshm05}
\end{figure}

To attempt to reproduce the NIRSPEC data, we generated a grid of models spanning
4200 K $\leq$ T$_{\rm eff}$ $\leq$ 4900 K, $-0.75$ $\leq$ [Fe/H] $\leq$ $+$0.50,
and $-1.00$ $\leq$ [C/Fe] $\leq$ $+$0.25. The resulting heat map for the sixth 
order, based on $\chi ^{\rm 2}$ fits to {\it only} the first two of the three 
strong lines in the wavelength interval 2.104 and 2.118 $\mu$m is shown in Fig. 
\ref{sshm02}. The best fit occurs for T = 4700 K, and [Fe/H] = $-$0.25. The fourth 
order contains a number of metal lines from Ti I and Fe I, though this spectral 
region is noisier than the data at shorter 
wavelengths. The resulting heat map, Fig. \ref{sshm04}, is consistent with the 
result for the sixth order, though the value of [Fe/H] is less constrained.
The results for the third order containing the $^{\rm 12}$CO$_{\rm (2,0)}$
bandhead is shown in Fig. \ref{sshm05}, and suggests [C/Fe] $\leq$ $-$0.5.

\begin{figure}[htb]
\centerline{{\includegraphics[width=15cm]{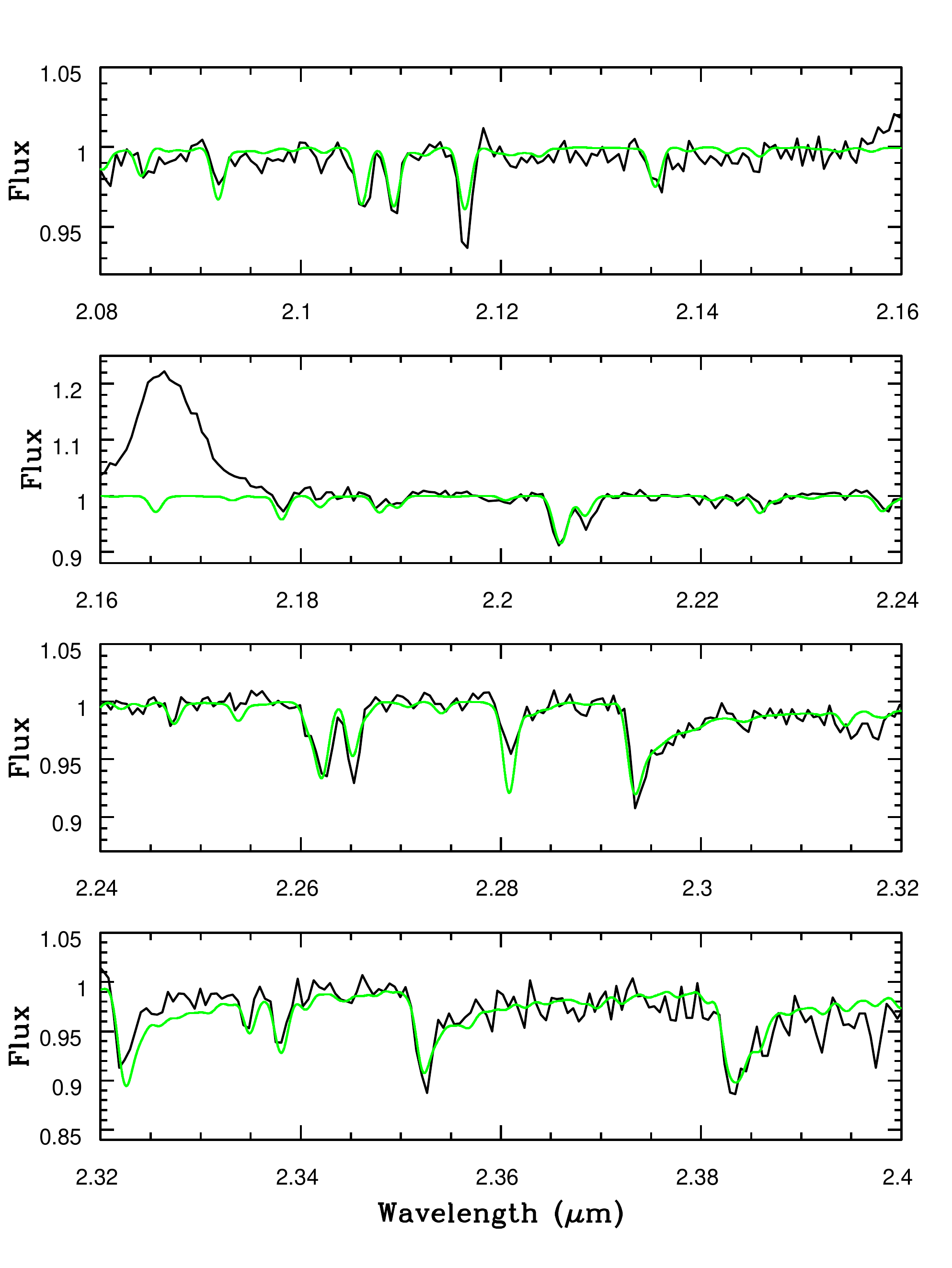}}}
\caption{The lower resolution IRTF SPEX spectrum of SS Cyg covering the
$K$-band from 2.08 $\mu$m to 2.4 $\mu$m with the best fit synthetic
spectrum derived for the NIRSPEC data plotted in green. The S/N ratio at 
2.20 $\mu$m was 115.}
\label{ssirtf}
\end{figure}

\begin{figure}[htb]
\centerline{{\includegraphics[width=15cm,angle=-90]{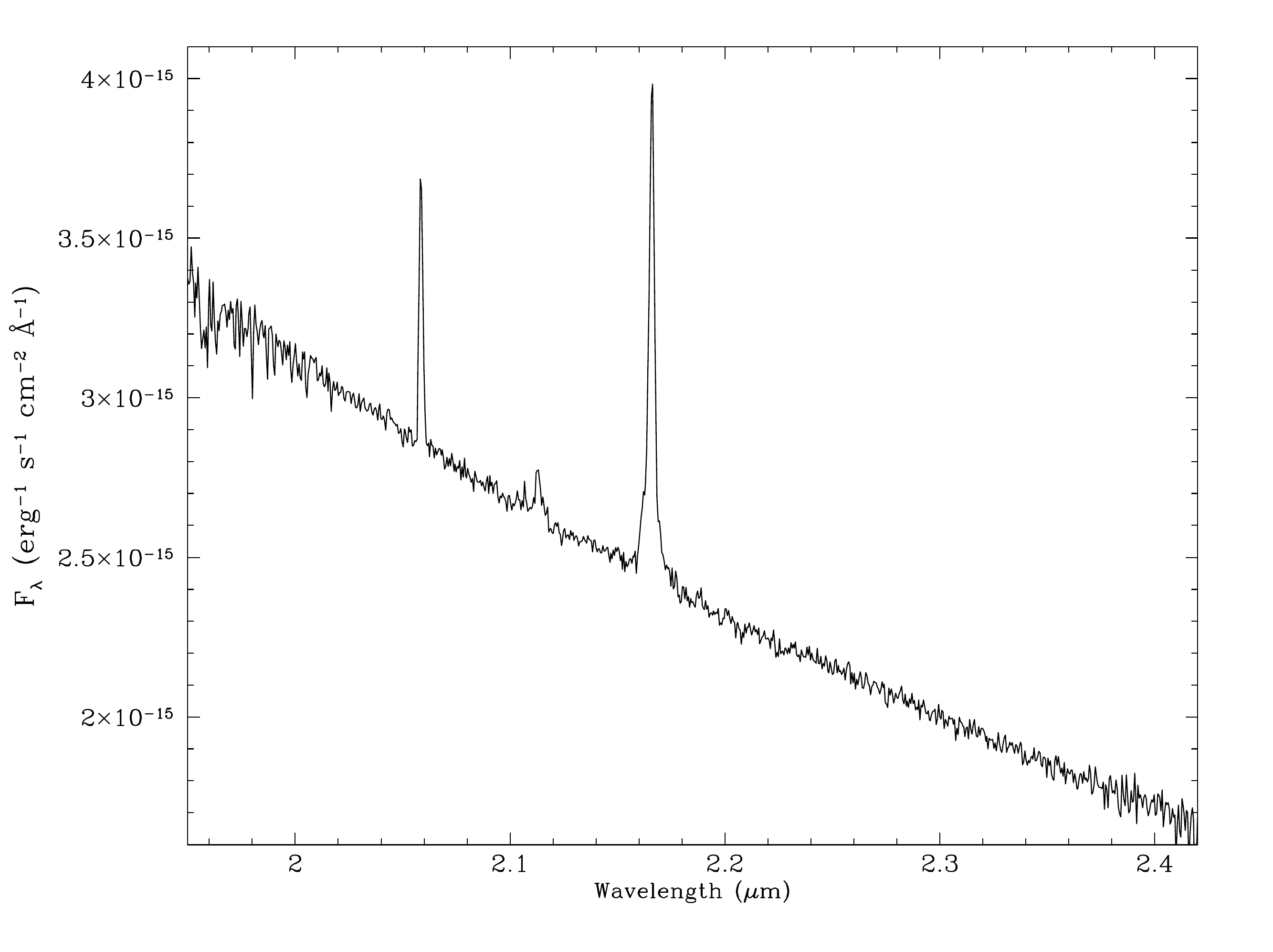}}}
\caption{The IRTF SPEX $K$-band spectrum of TT Ari, obtained on 2004
August 15. This disk dominated system has a weak He I emission feature
at 2.11 $\mu$m.}
\label{ttari}
\end{figure}

The best fit synthetic spectrum is overplotted on the data in Fig. 
\ref{sscyg} has these values, and fits the data set quite well. There is only 
one apparent anomaly in the NIRSPEC spectrum of SS Cyg, and that is the Al I 
line at 2.116 $\mu$m. We overplot the derived best fit 
model spectrum on the IRTF SPEX observation in Fig. \ref{ssirtf}. The fit is
quite good, given the S/N of those data. The main differences between the
model and the low resolution spectrum is again the Al I line 2.116 $\mu$m, and 
an Mg I feature at 2.282 $\mu$m. The Al I line was perfectly reproduced
for the solar spectrum and for $\epsilon$ Eri, but does appear to be slightly
under-predicted for the IRTF spectral sequence (though note our earlier 
discussion of the anomalous continuum in this bandpass). The line at 2.109 
$\mu$m is also due to Al I, thus the abundance of aluminum is not unusual. 
It is odd that this line is perfectly reproduced for our other CVs. There
is an He I line at this position that is often seen weakly in emission in the
$K$-band spectra of other CVs with more prominent accretion disks, such as TT 
Ari, see Fig. \ref{ttari}. This emission line might be the cause of the
peaks in the continuum on either side of this line in Fig. \ref{ssirtf},
but the effect of any such emission line would be to lessen the depth of the 
feature, not increase its strength.

The Mg I line at 2.282 $\mu$m is well reproduced by {\it kmoog} for all of the 
template star spectra. This line is very weak in the SPEX spectrum of SS Cyg. 
Since the Mg I line at 2.106 $\mu$m is at the appropriate strength in both the 
moderate and lower resolution data sets, we conclude that this is simply a S/N 
or telluric correction issue.

The NIRSPEC and SPEX data were obtained at quite different orbital phases,
0.20 and 0.40, respectively. The $K$-band light curve of SS Cyg presented in 
Harrison et al. (2007a) shows no evidence for irradiation at $\phi$ = 0.5,
and the $K$ magnitudes at these two phases in this light curve are nearly 
identical. Thus, the fact that the same synthetic spectrum works for both
epochs is not surprising. This does prove, however, that irradiation of
the secondary star is not the source that weakens the CO absorption
features, as the amount of the irradiated hemisphere seen at the dramatically
different orbital phases of our spectra has no influence on the strength
of the CO features.

\subsection{RU Peg}

RU Peg is a U Gem type dwarf nova of very long period (P$_{\rm orb}$ = 
8.99 hr), whose binary parameters were first examined by Kraft (1962), 
observing doubled Ca II lines at some points in the orbital phase and a 
spectrum broadly consistent with a G8 IV to K0 V. Friend et al. (1990) 
observed the $\lambda$8190\AA ~Na I doublet to measure radial velocity 
motion of RU Peg, finding that the secondary star is cooler and less massive 
than expected, with 
the spectrum being consistent with a K3 dwarf. They also note that the 
secondary star must be somewhat evolved, as the temperature, and masses 
derived are inconsistent with a main sequence dwarf. Harrison et al. (2004) 
presented IRTF observations of RU Peg, showing that the NIR spectrum was 
consistent with a spectral type of K2, but that RU Peg must contain a
subgiant since it would be over-luminous for a dwarf at the distance given by 
its parallax. Dunford et al. (2012) recently presented spectra of RU Peg 
analyzed with Roche tomograms, deriving a much cooler spectral type of K5. 

Along with the estimation of a cooler spectral type than found previously, 
Dunford et al. used moderate resolution visual band spectra to estimate 
$v_{\rm rot}$sin$i$ = 89 km s$^{\rm -1}$. As for SS Cyg, we convolved a solar
abundance model with T$_{\rm eff}$ = 4700 K to make our own estimate
of the rotation velocity and found a slightly higher value of 
$v_{\rm rot}$sin$i$ = 95 $\pm$ 10 km s$^{\rm -1}$. We have used this value
in generating the synthetic spectra that follow. Using the same size grid
as that for SS Cyg, the $\chi^{2}$ analysis for the sixth order,
shown in Fig. \ref{ruhm04}, suggests solar metallicities,
and 4600 $\leq$ T$_{\rm eff}$ $\leq$ 4700 K. The results for the fourth
order (Fig. \ref{ruhm02}) indicate slightly subsolar metallicities, and 4500 $\leq$ 
T$_{\rm eff}$ $\leq$ 4900 K. We conclude that the secondary star has 
T$_{\rm eff}$ = 4700 K, and [Fe/H] = 0.0. Analysis of the CO 
features in the third spectral order (Fig. \ref{ruhm05}) finds [C/Fe] = $-$0.75.

\begin{figure}[htb]
\centerline{{\includegraphics[width=15cm]{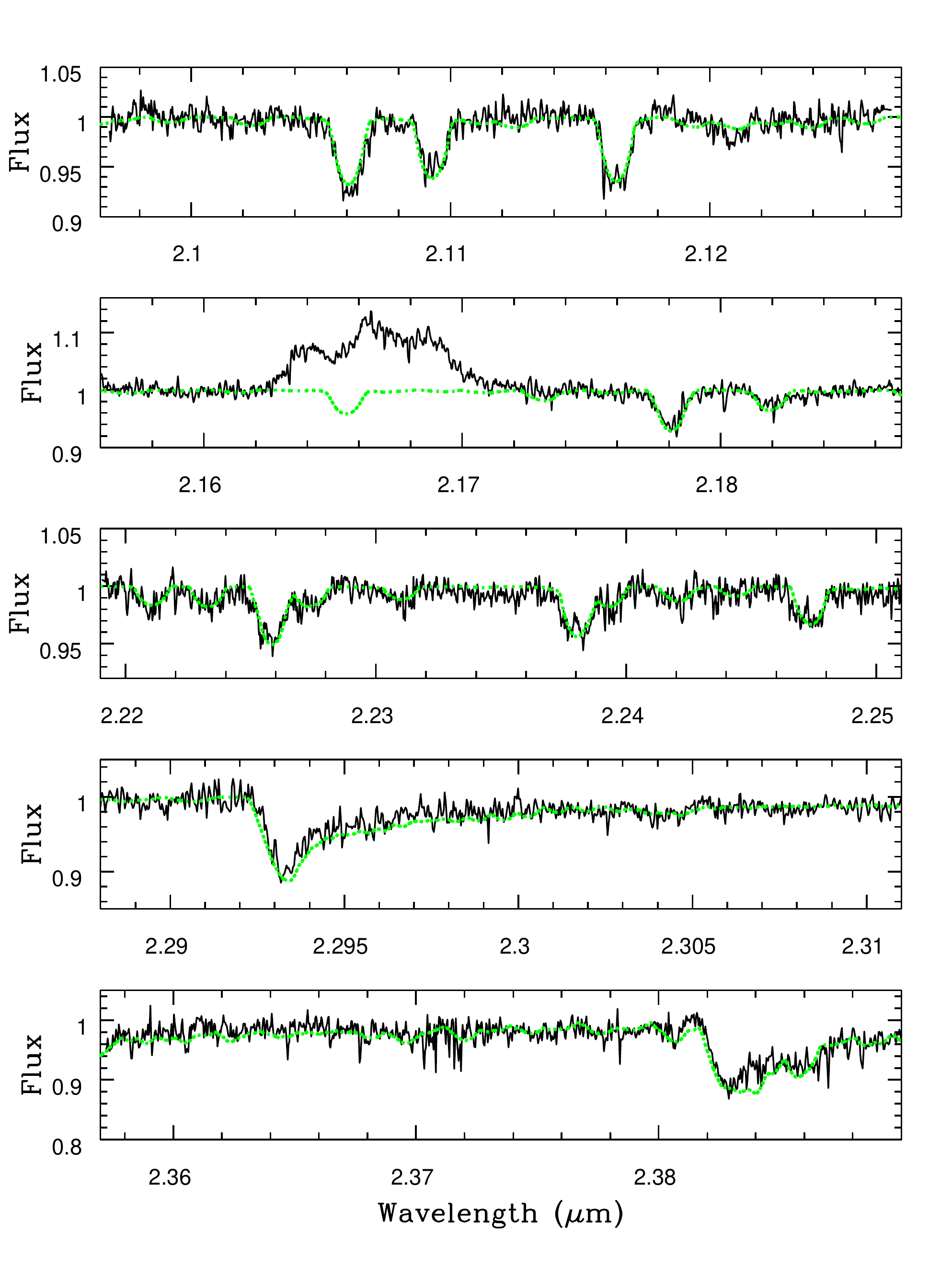}}}
\caption{The NIRSPEC spectrum of RU Peg (black) with the best fit synthetic
spectrum in green. The S/N ratio of these data ranged from 114 in the sixth 
order spectrum, to 69 in the second order spectrum.}
\label{rupeg}
\end{figure}

\begin{figure}[htb]
\centerline{{\includegraphics[width=15cm]{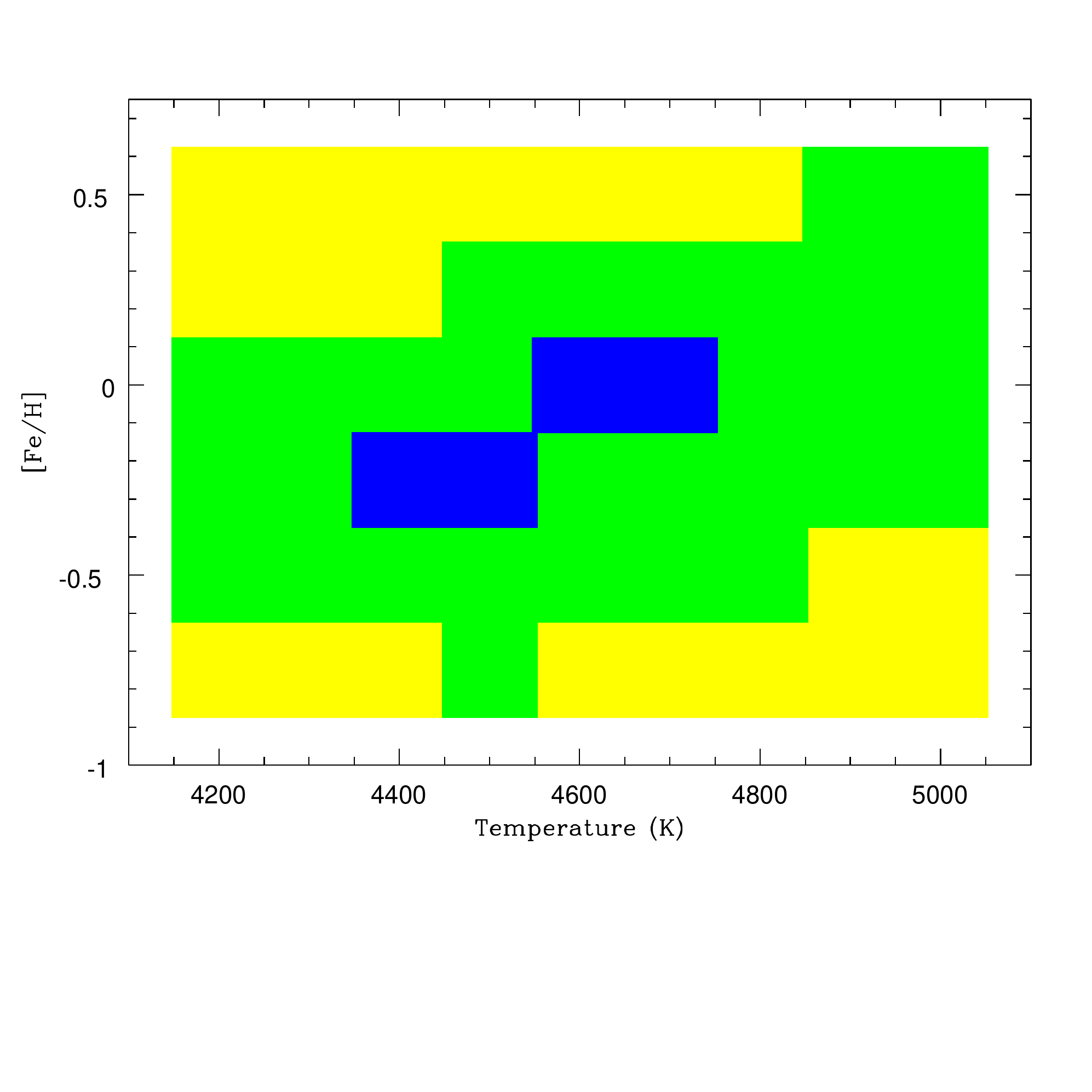}}}
\caption{The $\chi^{\rm 2}$ heat map for the model grid fits to the 
sixth order spectrum of RU Peg.}
\label{ruhm02}
\end{figure}

\begin{figure}[htb]
\centerline{{\includegraphics[width=15cm]{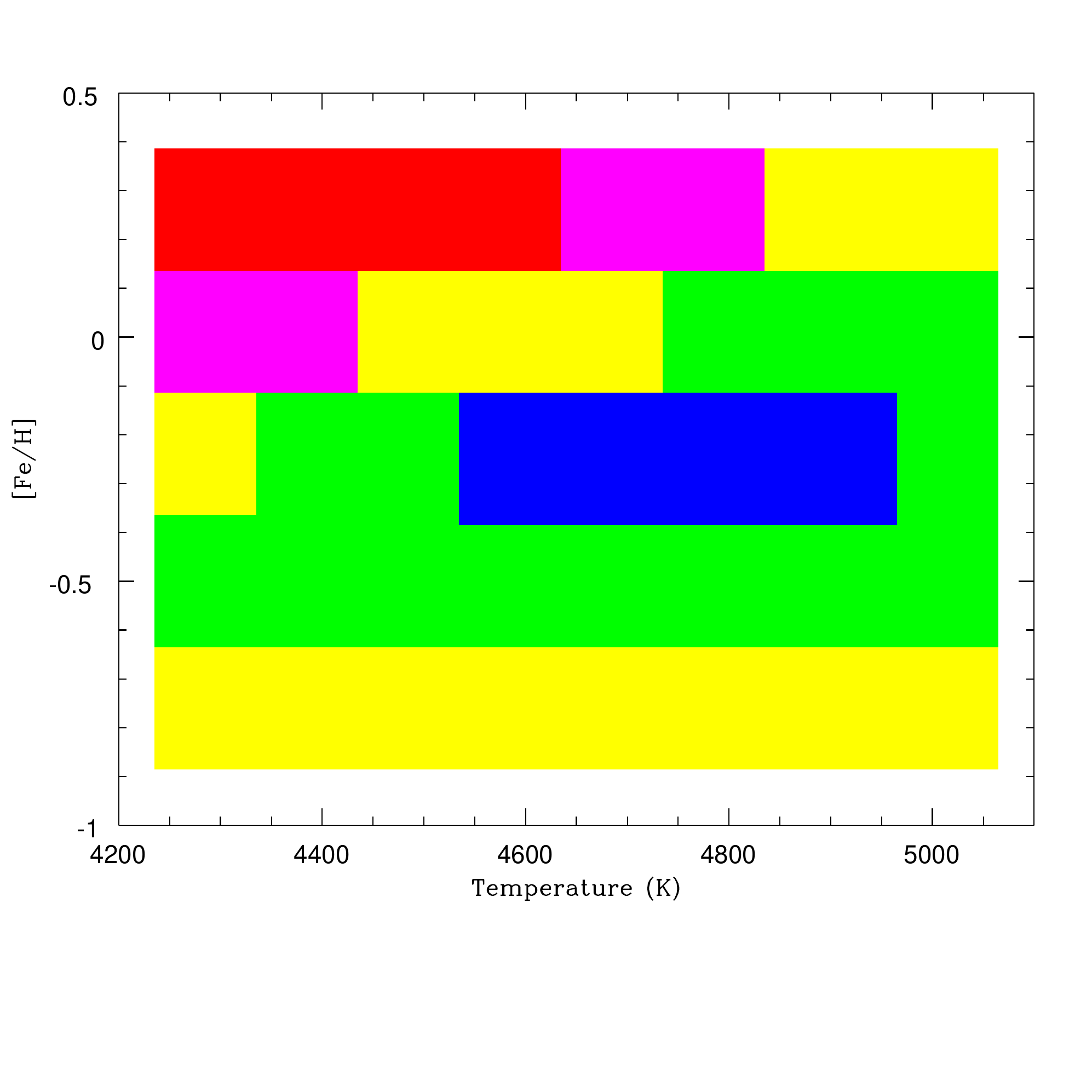}}}
\caption{The $\chi^{\rm 2}$ heat map for the model grid fits to the 
fourth order spectrum of RU Peg.}
\label{ruhm04}
\end{figure}

\begin{figure}[htb]
\centerline{{\includegraphics[width=15cm]{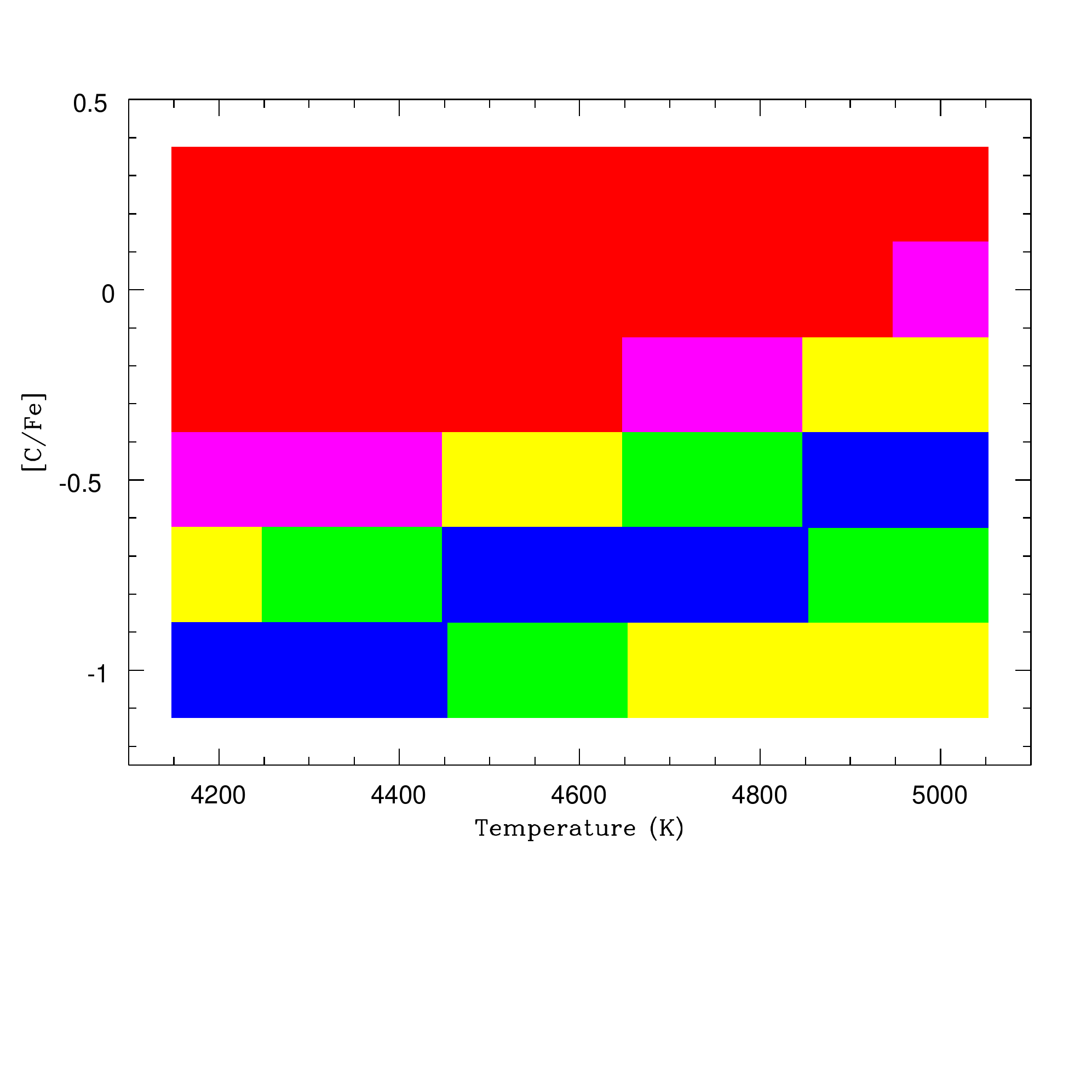}}}
\caption{The $\chi^{\rm 2}$ heat map for the model grid fits to the
third order spectrum of RU Peg. Like that for SS Cyg, a grid of single 
metallicity models (here with [Fe/H] = 0.0) were generated with only the 
carbon abundance altered for this analysis.}
\label{ruhm05}
\end{figure}

A synthetic spectrum with these values is overplotted on the NIRSPEC data
in Fig. \ref{rupeg}, and provides an excellent fit to the data in all of
the spectral orders. The H I emission from the disk is lower than seen in SS 
Cyg, and the underlying H I Br$\gamma$ absorption from the secondary star 
leaves its imprint on the emission line profile. Note that the long period 
and orbital phase at which the observations occurred make for minimal orbital 
smearing for the duration of our data taking: $\Delta v$ = 20 km s$^{\rm -1}$. 

\begin{figure}[htb]
\centerline{{\includegraphics[width=13cm]{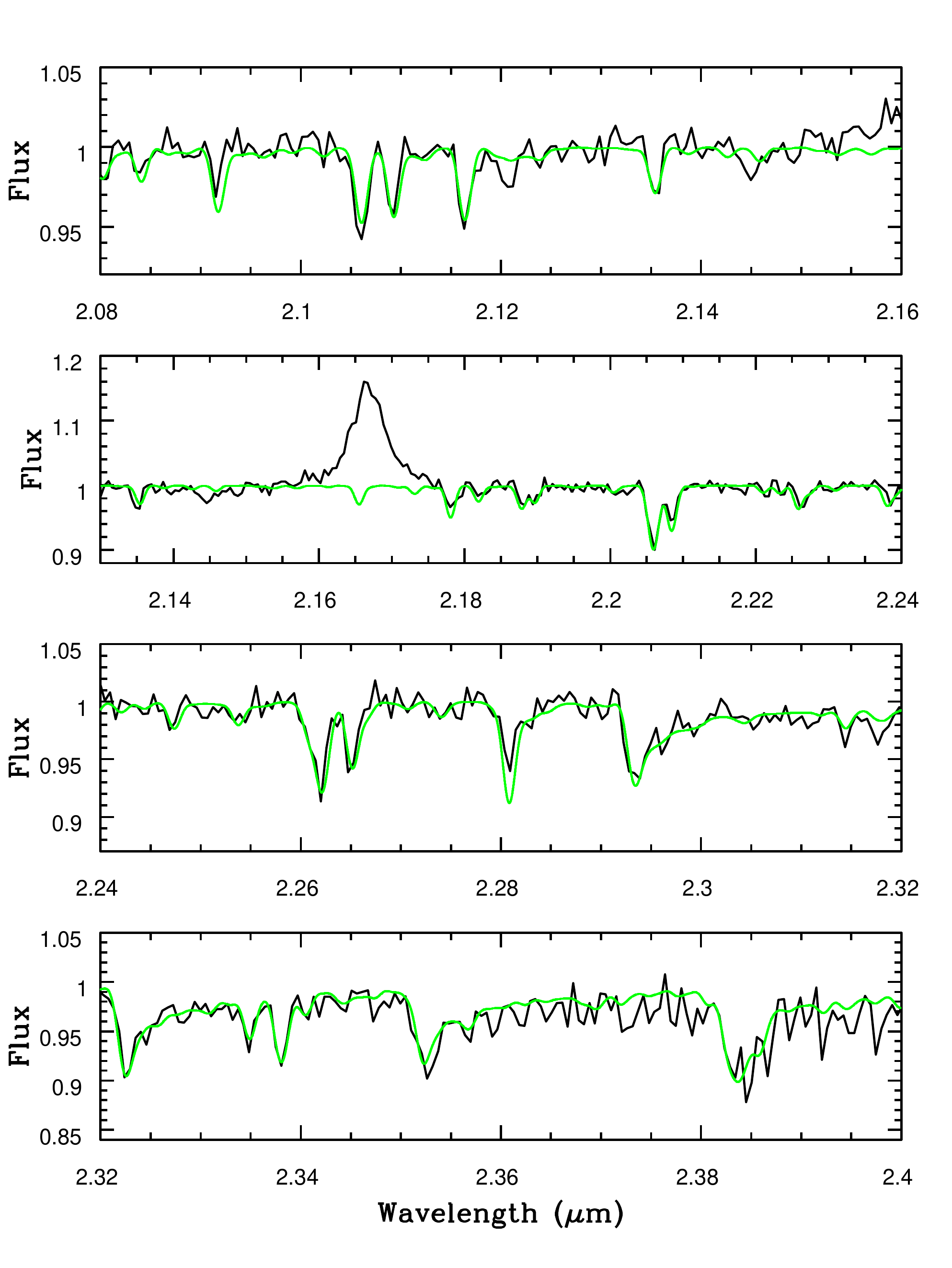}}}
\caption{The SPEX spectrum of RU Peg (black) with the best fit synthetic
spectrum plotted in green. The S/N ratio at 2.20 $\mu$m was 114.}
\label{rupegirtf}
\end{figure}

The final model derived for the NIRSPEC spectrum is overplotted on the SPEX
data in Fig. \ref{rupegirtf}. This model fits the lower resolution data
very well, including all of the other strong $^{\rm 12}$CO bandheads in the 
red part of the $K$-band. Sion et al. (2004) found that the carbon abundance 
in the photosphere of the white dwarf was 10\% of solar. Thus, the carbon 
deficit observed in the UV can be directly traced to the secondary star. Sion 
et al. also found that silicon suffered a similar deficit as that of carbon. 
The most prominent Si I lines in the wavelength range plotted in Fig. 
\ref{rupegirtf} are at 2.136 and 2.189 $\mu$m, and while these lines are weak 
in our data, and the S/N not very high, they seem to be well represented by 
our solar abundance model.

\subsection{GK Per}
GK Per is an old classical nova, having experienced an extremely luminous 
nova eruption in 1901 (see Harrison et al. 2013). It has since exhibited 
periodic dwarf nova outbursts every 2-3 years since 1966, brightening by 
$\sim$ 3 mag during these events. Kraft (1964) detected the 
mass donor star in the optical, adopting a spectral type of K2 IV and noting 
that the spectrum was highly variable with an equivalent width of Sr II 4077 
\AA\ consistent with a giant at one point and a dwarf the next. More recently,
Morales-Reuda et al. (2012) analyzed spectra of GK Per, finding that the 
donor star appeared as K1 IV. Their analysis consisted of an iterative 
template subtraction method, however, not direct detection of the secondary 
star. Harrison et al. (2007b) presented a $K$-band spectrum of the mass donor 
star in GK Per, and found that it has extremely weak CO features.

The radial velocity study conducted by Morales-Reuda et al. found
$v_{\rm rot}$sin$i$ = 61.5 $\pm$ 11.8 km s$^{\rm -1}$, 
K$_{\rm 2}$ = 120.5 $\pm$ 0.7 km s$^{\rm -1}$, and a mass ratio of $q$ = 0.55
$\pm$ 0.21. Given the luminosity of its classical nova eruption suggests a 
massive white dwarf. If this is true, the radial velocity solution implies 
a surface gravity for the secondary star of log $g$ $\leq$ 4. As noted above, 
it is not possible to constrain the surface gravity of the program objects 
given the limitations of our data set. Thus, in contrast to SS Cyg and RU Peg, 
we set the surface gravity to log $g$ = 4.0 for our modeling of GK Per (though 
this produced identical results to models with log $g$ = 4.5). As for the 
preceding two objects, we analyzed the rotational velocity apparent in our 
NIRSPEC data and find $v_{\rm rot}$sin$i$ = 55 $\pm$ 10 km s$^{\rm -1}$, 
slightly lower than that derived by Morales-Rueda et al., but agreeing within 
the error bars of the two measurements. We use this value in all of our 
modeling.

\begin{figure}[htb]
\centerline{{\includegraphics[width=13cm]{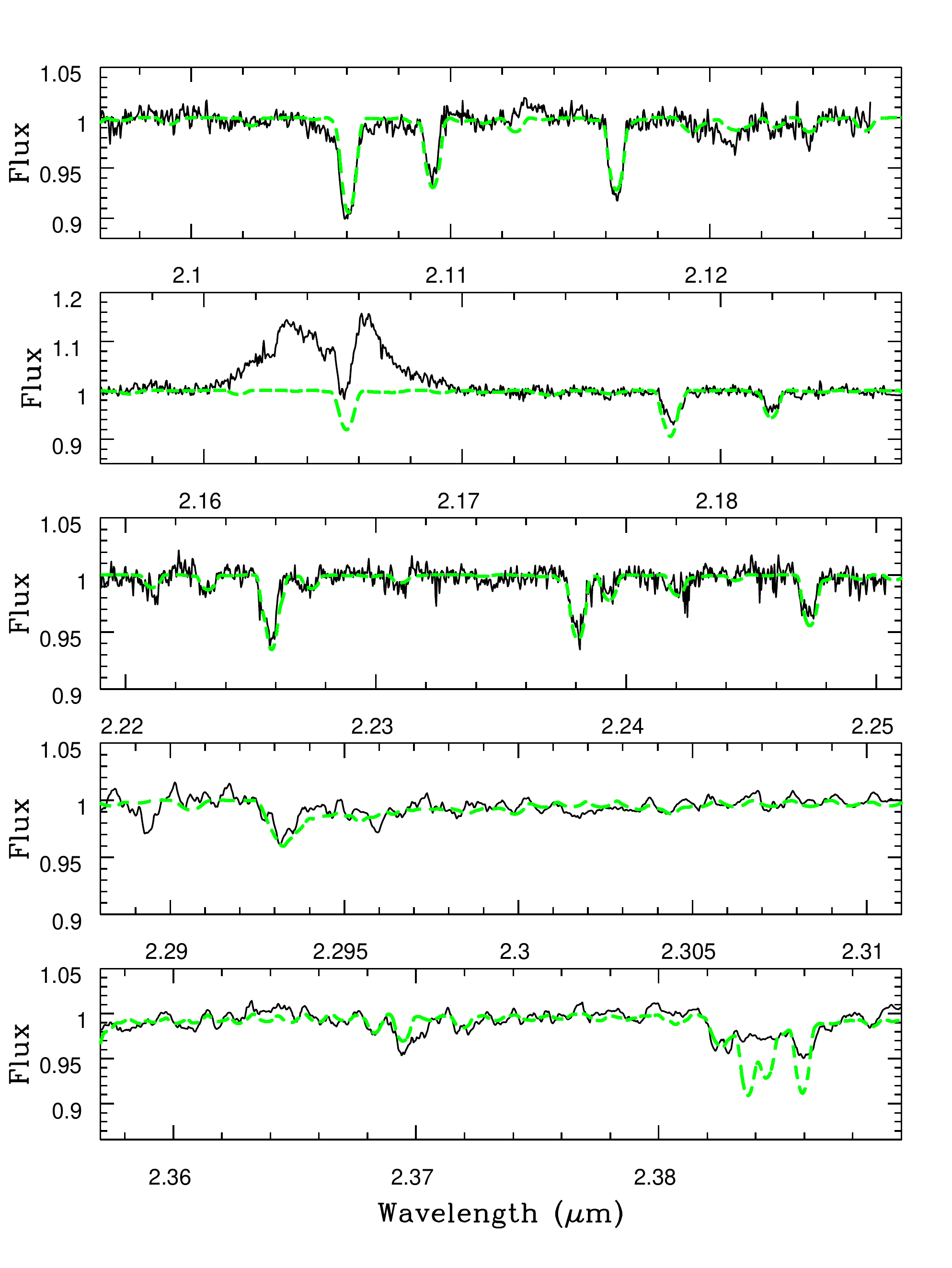}}}
\caption{The NIRSPEC spectrum of GK Per (black). We have boxcar smoothed
the second and third order spectra (by 9 pixels) to improve their S/N ratios. 
The final, best fit synthetic spectrum is plotted in green. The S/N ratio
of these data ranged from 150 in the sixth order, to 50 in the second order.}
\label{gkper}
\end{figure}

\begin{figure}[htb]
\centerline{{\includegraphics[width=13cm]{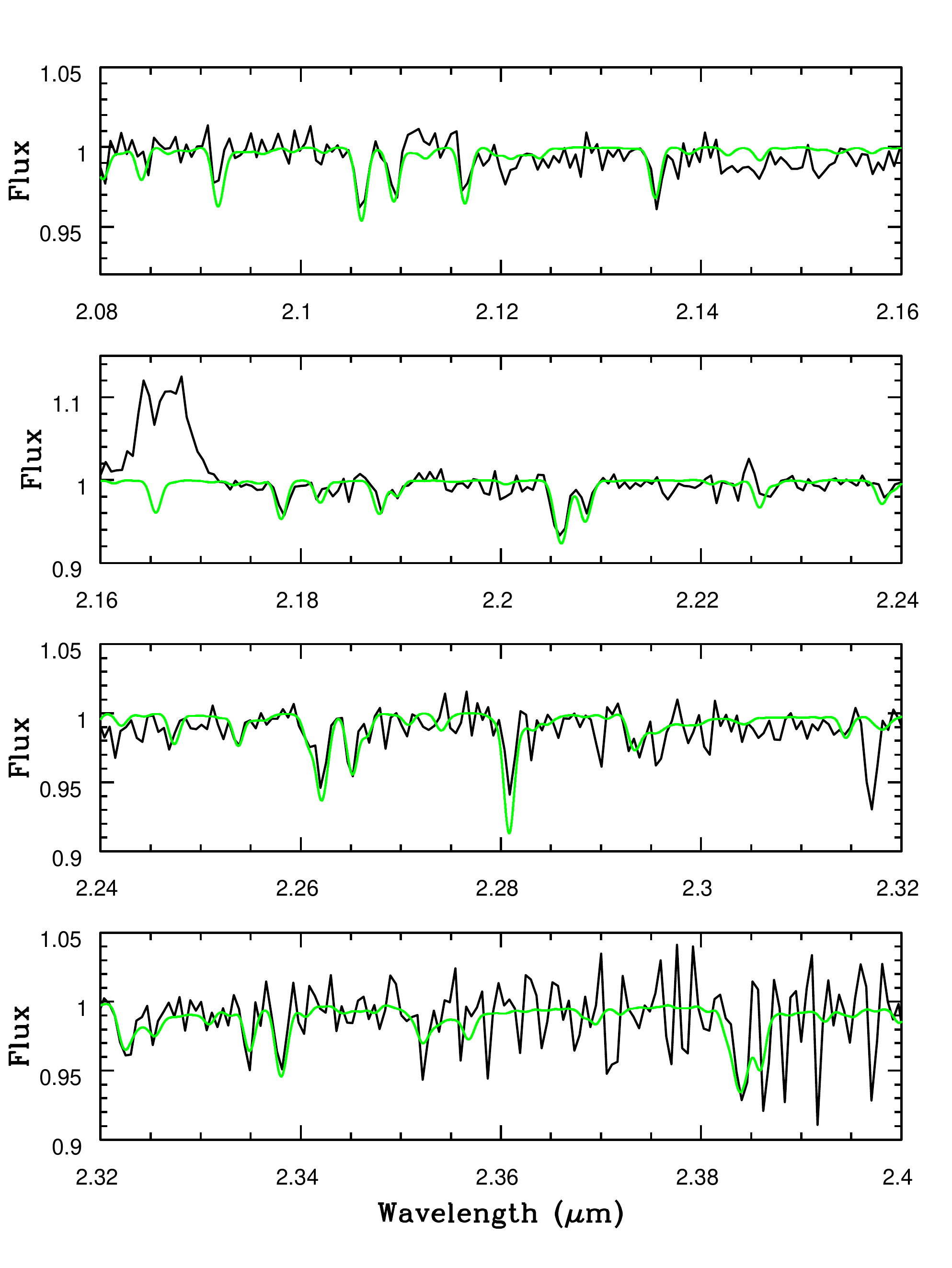}}}
\caption{The SPEX spectrum of GK Per (black) with the best fit synthetic 
spectrum plotted in green. The S/N ratio at 2.20 $\mu$m was 91.}
\label{gkperirtf}
\end{figure}

The NIRSPEC data is plotted in Fig. \ref{gkper}. $\chi^{\rm 2}$ analysis of
the model grid found these best fit values: T$_{\rm eff}$ = 5100 K, 
[Fe/H] = $-$0.125, and [C/Fe] = $-$1.0. GK Per is fainter than our other
two targets so the CO region is very noisy, and thus the measurement of 
[C/Fe] for GK Per is slightly less certain than found for the other two
targets. It is interesting to note that the H I absorption of the underlying
secondary star is sufficient to create the appearance of no H I Br$\gamma$ 
emission at the center of this feature. The SPEX spectrum of GK Per, 
Fig. \ref{gkperirtf}, is even noisier than the NIRSPEC data. Clearly, however, 
the same model does an excellent job at fitting the observations. 

\subsection{Adopted Errors in Parameters}
\begin{figure}[htb]
\centerline{{\includegraphics[width=13cm]{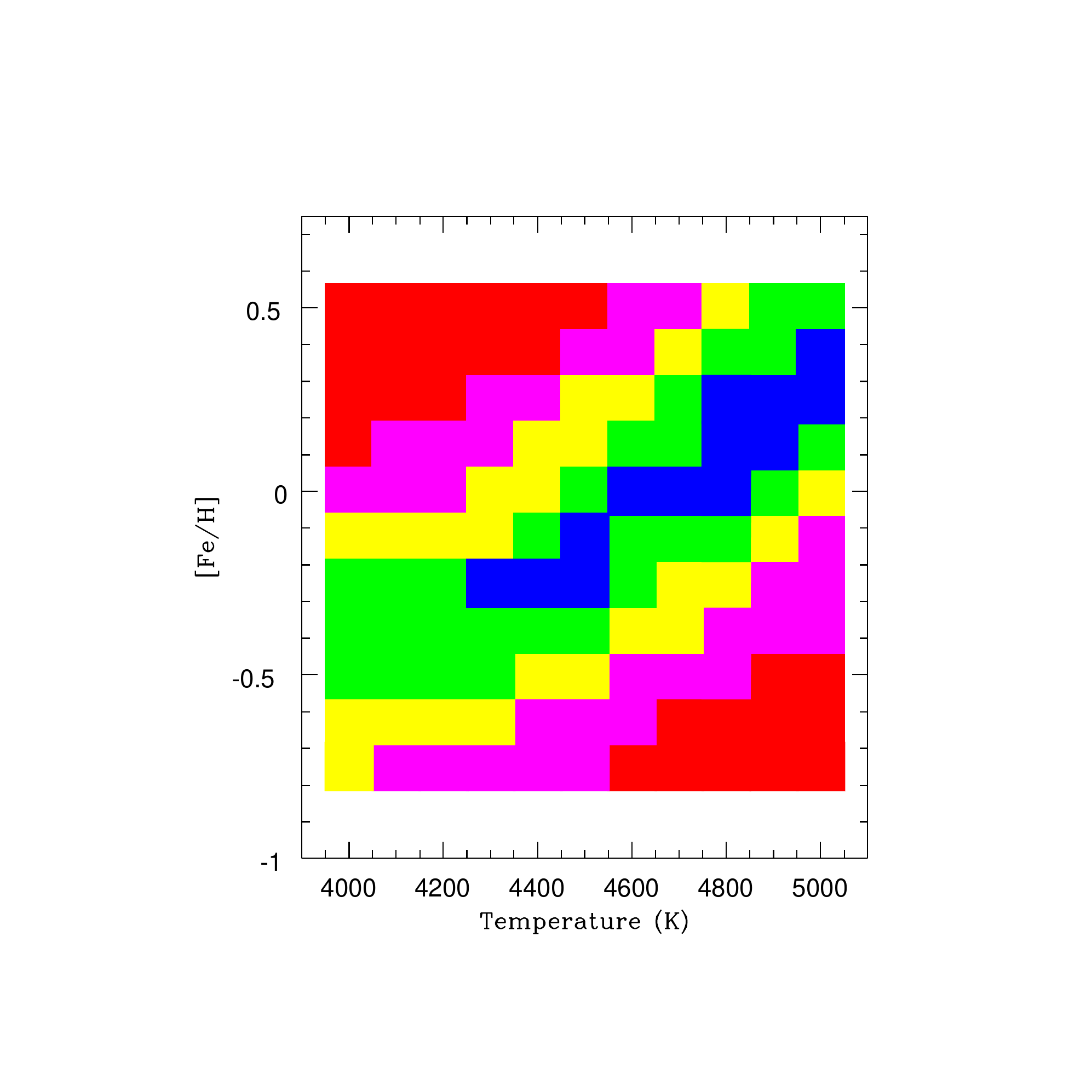}}}
\caption{A heat map for the $\chi^{\rm 2}$ analysis of the IRTF spectral
template HD36003 for a large grid of model spectra. The mean of the
published values for this star, see Table 2, are T$_{\rm eff}$ = 4550 K,
and [Fe/H] = $-$0.17. The grid has temperature intervals of 100 K, and
[Fe/H] stepped by $\pm$0.125.}
\label{k4vhm}
\end{figure}

As demonstrated above, we are easily able to reproduce the spectra of
objects with known parameters. Our examination of K stars in the IRTF spectral 
library shows that if we know the value of [Fe/H], we can derive values
of T$_{\rm eff}$, or vice versa, that are correct to within the published 
error bars for that object.  To examine how well we can determine T$_{\rm eff}$ 
or [Fe/H] if we do not know either value, we present a $\chi^{\rm 2}$ analysis 
for the IRTF template HD36003. The heat map for the spectral region 2.18 
$\leq$ $\lambda$ $\leq$ 2.37 $\mu$m of this star is shown in Fig. \ref{k4vhm}. 
While the models with parameters similar to the published values for both 
T$_{\rm eff}$ (4615 $\pm$ 29 K) and [Fe/H] ($-$0.14 $\pm$ 0.08) do have low 
values of $\chi^{\rm 2}$, there is a range of solutions with quantitatively 
similar fits to the data. As demonstrated in the plots shown in
Figs. \ref{k0tok3} and \ref{k4tok7}, for K dwarfs, the strong Na I doublet
(at 2.209 $\mu$m), the Ca I triplet (at 2.26 $\mu$m), and the CO features all 
slowly get stronger as we decrease temperature. Thus, models can mimic the 
observed  spectrum by either an enhanced abundance of [Fe/H] and a hotter 
temperature, or a reduced value of [Fe/H] and a lower temperature.

Our $\chi^{\rm 2}$ analysis over such a large span in wavelength is certainly 
also affected by missing transitions for weak lines, the S/N of the data, and 
for this lower resolution spectrum, the continuum subtraction process. The red 
end of the $K$-band is filled 
with strong telluric features, and even for high S/N data, small discrepancies 
can appear due to a slight mis-match of the telluric standard with the program 
object. This is especially clear for the CO features, as some bandheads in a 
particular spectrum are well fitted by the models, while others are not.

Given these results, if we are able to constrain the limits on the temperature
of the object, such as ``the secondary star in SS Cyg has a spectral type
near K5'', and not expect super-solar metallicities, we find that
temperatures derived from these data are good to $\pm$ 250 K, and [Fe/H] is
good to $\pm$ 0.25. Moderate resolution data that covered either the Na I
doublet or Ca I triplet would almost certainly lead to more precise values
of both quantities.

Putting error bars on our values for [C/Fe] is more difficult in that none of
our templates are known to have unusual values for this parameter. It is clear 
that our models reproduce the CO features in the template spectra over a large 
range in metallicity. Thus, altering just the abundance of carbon should
not introduce any significant issues, as it simply acts to reduce the strength
of the CO features to mimic those seen in lower metallicity stars. How
a reduced carbon abundance affects the input stellar atmosphere is difficult
to assess given the unknown evolutionary state of the CV secondary stars. The
presence of more than one CO bandhead in all of the spectra also allows us
to have more confidence in a particular value of [C/Fe]. For the higher
S/N spectra, we expect that the precision of the values for [C/Fe] are similar
in scale to [Fe/H]: $\pm$ 0.25. For GK Per the data are quite poor compared
to the other two CVs, and the derived value for [C/Fe] is less certain. It is 
obvious, however, that the value of [C/Fe] is lower in GK Per than in the
other two sources. This result will be amplified if the gravity of GK Per is 
significantly lower than our input value of log $g$ = 4.0.

\section{Discussion}

We have obtained moderate resolution $K$-band spectroscopy of three CVs
that were previously noted to have weaker than expected CO features: SS
Cyg, RU Peg, and GK Per. Because those earlier data were obtained at
lower resolution, R $\sim$ 2,000, questions were raised about the possibility
of ``veiling'' of the CO features to make them appear weaker. While we
show below that this argument has no validity if realistic veiling scenarios
are examined, it is the increased dispersion of the spectra obtained with
NIRSPEC that easily argues against such a scenario. The NIRSPEC data have
a dispersion of 0.32 \AA/pix, nearly 1/17 that of the SPEX data. Thus, 
the flux per pixel from any veiling component would be diminished by this 
factor. Model spectra fitted to the two different sets of data would give 
different values for the strength of the CO features if there was some type 
of veiling confined to the CO region.

To provide quantitative estimates of the relative carbon abundances for CVs we
have altered MOOG to allow us to model the entire $K$-band, and developed a 
software package that constructs large grids of synthetic spectra, and then 
searches these grids to find the best fitting solution to the observed data. We
specifically allowed for the abundance of carbon to be a separate input
under the assumption that the $^{\rm 12}$C abundance has been changed in
the initial branches of the CNO cycle, as argued in Harrison et al. (2004). 
Obviously, weak CO features could be explained by a deficit of oxygen, but a 
process that could cause such a deficit is not obvious. With this 
software we derived values of T$_{\rm eff}$, [Fe/H], and [C/Fe] from the 
NIRSPEC spectra for the three program objects. We list these results in 
Table 3. Except for the deficits of carbon, no other abundance anomalies
are apparent, and the secondary stars of all three objects have solar, or
slightly sub-solar abundances.

The total $^{\rm 12}$C abundance for SS Cyg and RU Peg are similar,
$\sim$ 25\% of solar, while for GK Per it is $\sim$ 10\% of solar. If these 
peculiar values are the result of the CNO cycle (c.f., Marks \& Sarna 1998), 
than the abundance of $^{\rm 13}$C should be enhanced, allowing for the 
detection of the bandheads 
of $^{\rm 13}$CO. The two strongest such bandheads with $\lambda < $ 2.4 $\mu$m
are $^{\rm 13}$CO$_{\rm (2,0)}$ at 2.345 $\mu$m, and 
$^{\rm 13}$CO$_{\rm (3,1)}$ at 2.374 $\mu$m. The first of these is not covered 
by our NIRSPEC data, but Dhillon et al. (2002) present a $K$-band spectrum of 
SS Cyg which appears to show a strong $^{\rm 13}$CO$_{\rm (2,0)}$ absorption 
feature. They attribute this feature to metal line absorption in star spots 
as indicated by anomalously strong Na I absorption features (at 2.3355 and 
2.3386 $\mu$m) in their spectra. These two Na I lines are not anomalous in our 
SPEX data, nor is there a $^{\rm 13}$CO$_{\rm (2,0)}$ feature of the strength 
observed by Dhillon et al. The second of the $^{\rm 13}$CO features is covered 
by our NIRSPEC observations, but there is a deep telluric feature at this 
wavelength that makes good correction quite difficult, and results in a noisy 
portion of the second order spectrum. Even with this issue, there is certainly 
no sign of a strong absorption feature at this location in either SS Cyg or RU 
Peg. The SPEX data covers both features, and while there are hints of both of 
these bandheads in the data for SS Cyg and RU Peg, the spectra are simply too 
noisy to draw any firm conclusions. New, higher S/N moderate resolution 
spectroscopy will be required to properly investigate the 
$^{\rm 12}$C/$^{\rm 13}$C ratio.

\subsection{A Closer Inspection of the Continuum of SS Cyg, and an Examination
of Realistic Veiling Scenarios}

\begin{figure}[htb]
\centerline{{\includegraphics[width=13cm]{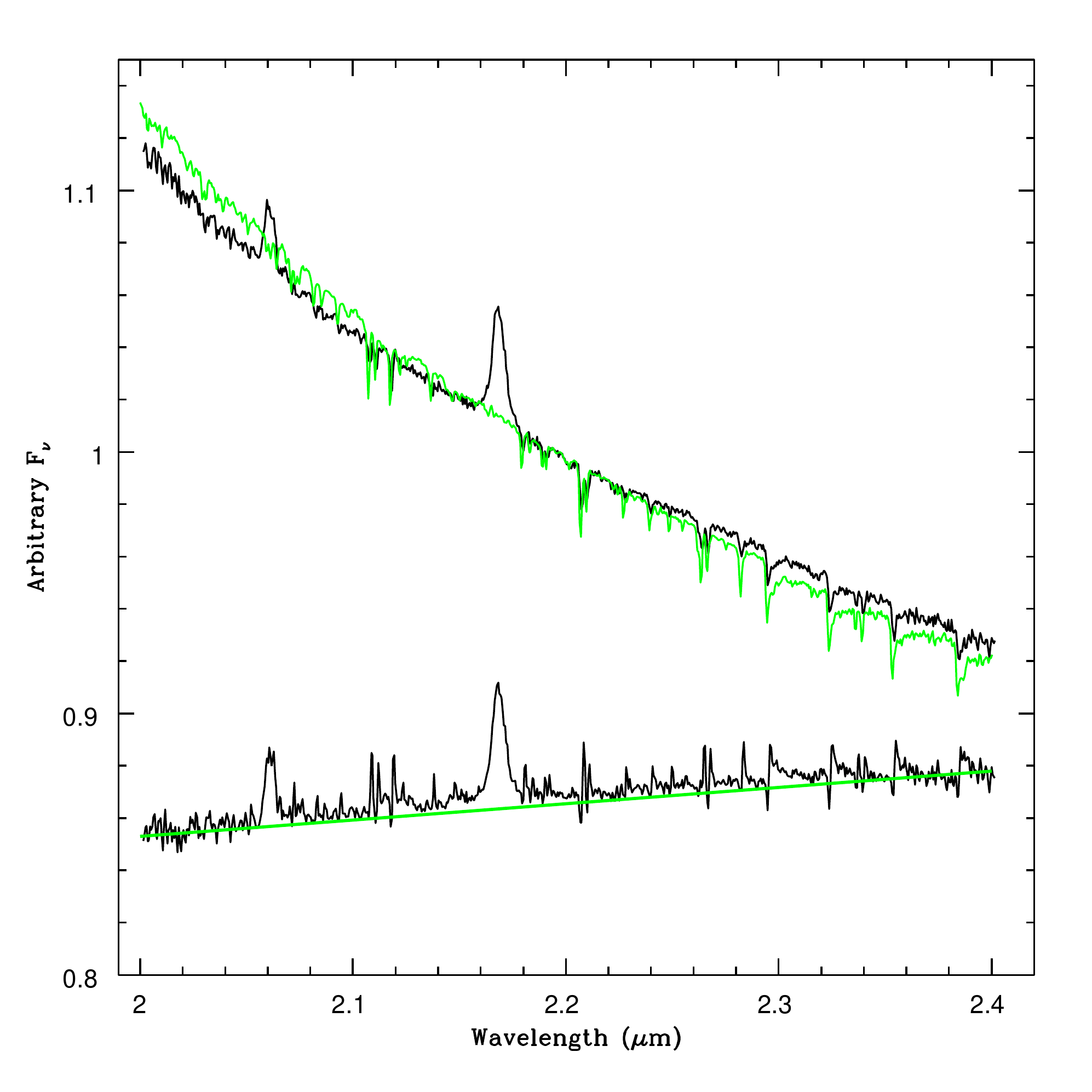}}}
\caption{The SPEX spectra of SS Cyg (black) and the K4V template HD45977 (green)
normalized at 2.2 $\mu$m. Subtraction of the spectrum of HD45977 from that of SS Cyg
leaves the residual spectrum (bottom) that has a flat slope (f$_{\rm \nu}$
$\propto$ $\nu^{\rm -0.16}$).  }
\label{sscygcomp}
\end{figure}

Harrison et al. (2004) found that the $K$-band continua of many long
period CVs were flatter than expected if the secondary star was the sole
source of emission in that bandpass. To more closely examine the slope
of the continuum of SS Cyg, we compare the full $K$-band spectrum
of SS Cyg with that of the IRTF template star that has the most similar 
temperature: HD45977. In Fig. \ref{sscygcomp}, we subtract the spectrum of 
HD45977 from that of SS Cyg, both spectra having been normalized at 
2.2 $\mu$m. The result is a residual spectrum that is fairly flat,
with f$_{\nu}$ $\propto$ $\nu^{\rm -0.16}$. Given the uncertainties in
the temperatures of the two objects, and the differing sources of the
data, such a flat spectrum is consistent 
with free-free emission. The continuum subtraction process step required to
allow us to directly compare the SPEX data to the synthetic spectra easily
removes a weak continuum source such as this, and this is why the results for
the NIRSPEC and SPEX are simultaneously consistent with the same model.

\begin{figure}[htb]
\centerline{{\includegraphics[width=13cm]{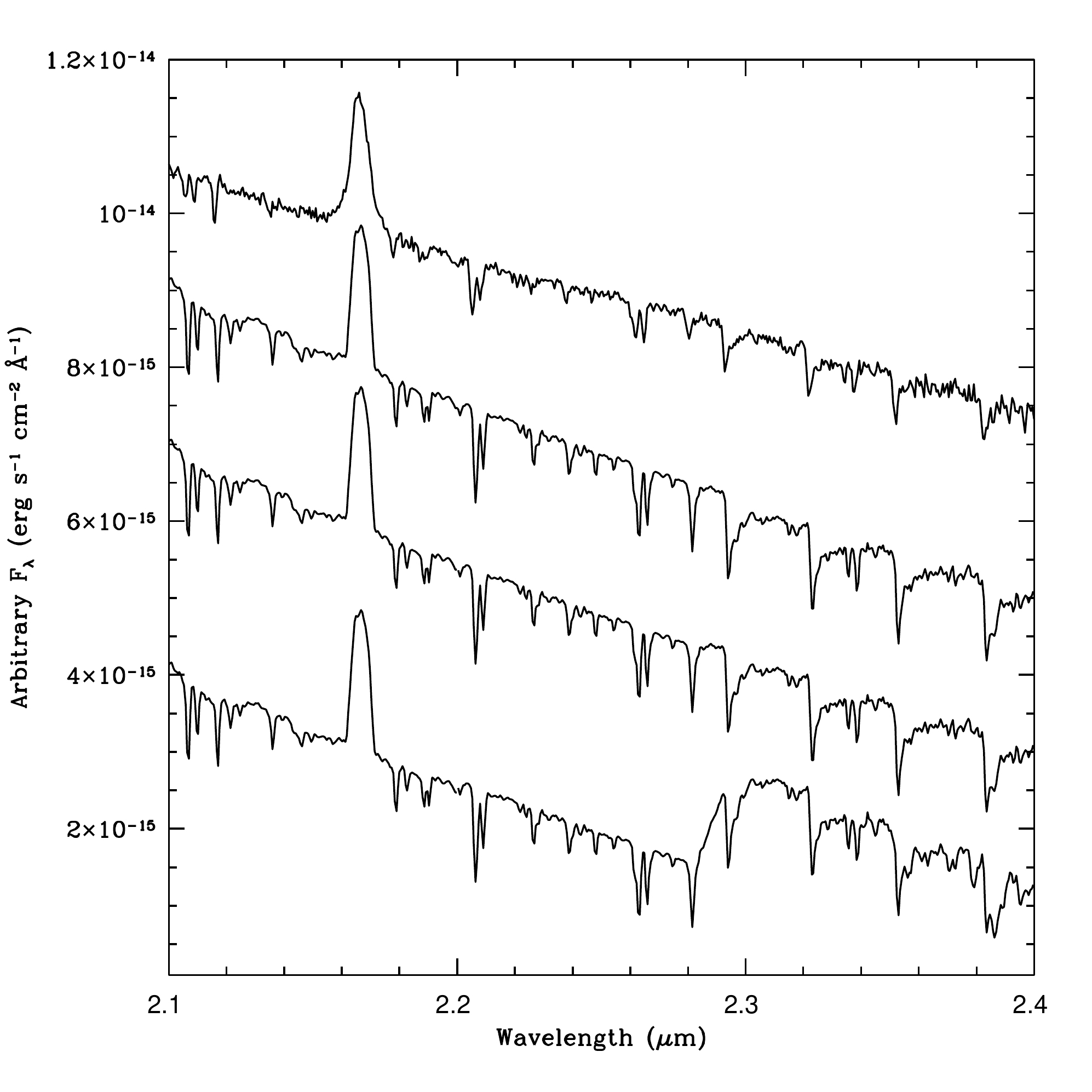}}}
\caption{The SPEX spectrum of SS Cyg (top). The next spectrum down
is a synthetic spectrum for a solar abundance K4V to which we have 
we have added a Gaussian broadened H I emission line spectrum to match
the profile and strength of the H I Br $\gamma$ line of SS Cyg. The spectrum
below that has had the emissivities of the Pfund lines increased by a factor
of 10. The bottom spectrum has the Pfund line emissivity increased by a
factor of 100.}
\label{sscygpfcomp}
\end{figure}

We can also attempt to model the contribution of line emission sources that
might preferentially veil the CO features. There are only two possible
sources that we can envision that might do this: the H I Pfund continuum,
and CO emission. As shown in Fig. \ref{sscygcomp}, the H I Br$\gamma$ emission line 
in SS Cyg is quite strong, so the Pfund series of hydrogen will also be present.
The Pfund series limit is at $\lambda$2.279 $\mu$m, and the transitions
Pf23 and higher, will fall in our modeling bandpass. The emissivities of
these lines are very small; Hummer \& Storey (1987) list them as
being $\leq$ 1.7\% that of Br$\gamma$ (for ``Case B'' conditions). Fitting
the Br$\gamma$ line with a Gaussian profile gives FWHM = 73 \AA. We will assume
all of the H I lines have this profile. In Fig. \ref{sscygpfcomp}, we present 
spectra where we added a rotationally broadened spectrum of the Pfund series 
(up to Pf100) to the K4V spectral template, and compare it to the spectrum of 
SS Cyg. Given their low emissivities, the effect of the H I lines is not 
detectable. We also artificially multiplied the emissivities of the Pfund 
lines by factors 10 and 100 and added them to the K4V template. Only at
very large multiplicative factors can their presence be detected, and
even then, they add a smooth continuum to the CO region, leaving the absorption
features relatively undisturbed, but producing a highly distorted continuum. 
Even with this simplistic model, it is clear that H I Pfund emission cannot 
provide the veiling necessary to explain weak CO features.

The detection of CO emission in WZ Sge (Howell et al. 2004) has led to the
suggestion that CO emission is the source of the veiling needed
to produce weaker than expected CO features seen in the program CVs. Since the 
CO emission must come from the accretion disk, a similar result as that for 
the Pfund continuum emission attains. As discussed in Scoville et al. (1980), 
such emission is expected to be confined to regions where T $>$ 3000 K, 
and where the densities are high, $n_{\rm e}$ $\geq$ 10$^{\rm 10}$ 
cm$^{\rm -3}$.  Since SS Cyg was in a minimum during our observations, 
presumably these conditions exist somewhere within its accretion disk. 
As shown in Bik et al. (2006), optically thin CO emission features are 
essentially the inverse of the CO absorption features seen in late type stars. 
This similarity allows us to construct crude models to investigate what CO 
emission might look like if its origin was in a CV accretion disk. 

\begin{figure}[htb]
\centerline{{\includegraphics[width=17cm]{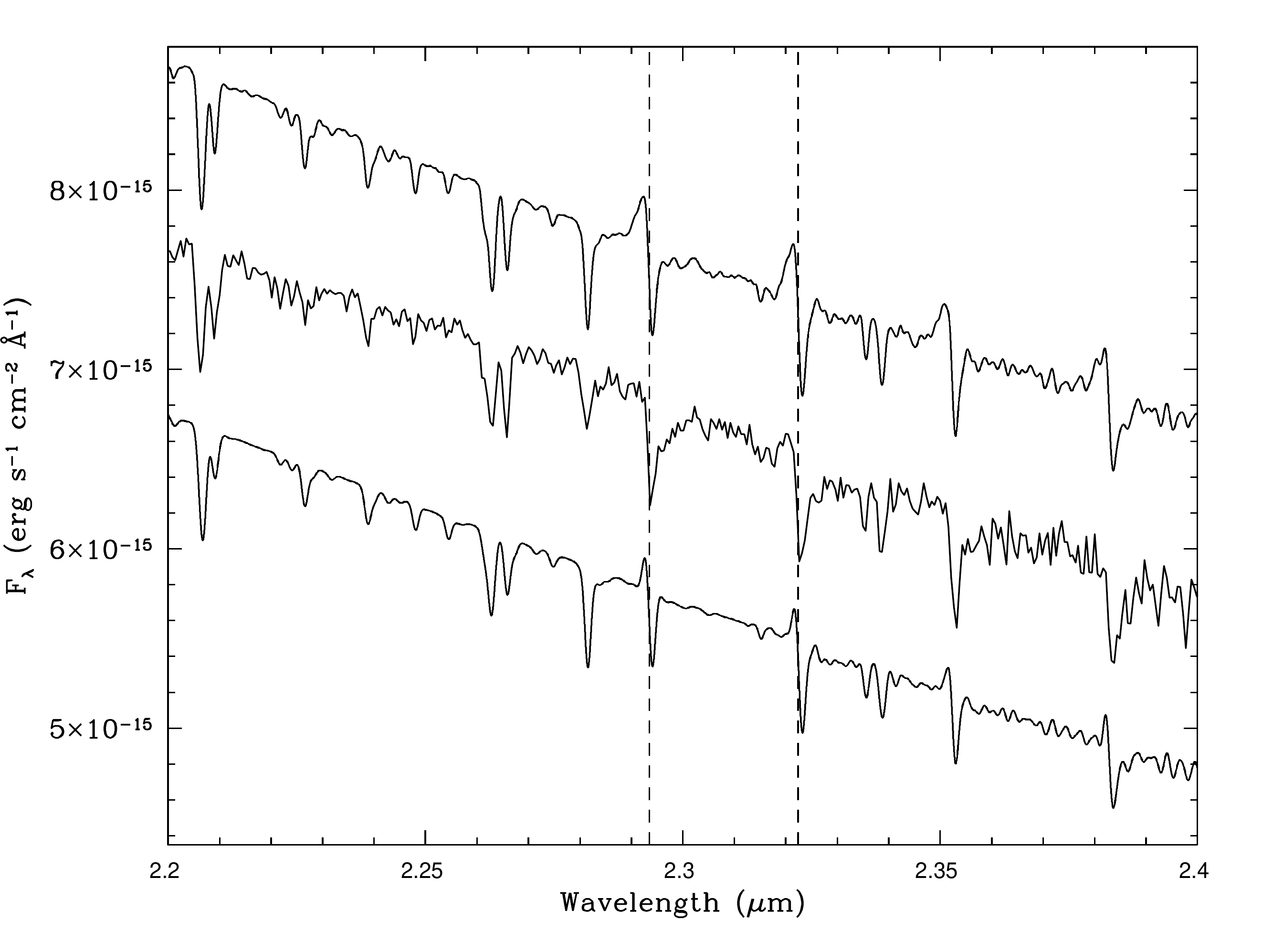}}}
\caption{The SPEX spectrum of SS Cyg (middle). The top spectrum results
from the addition of a CO emission line spectrum Gaussian broadened to
match the Br $\gamma$ line, added to the best fitting (solar abundance)
synthetic spectrum. The bottom spectrum is the result of the identical 
process, except the velocity broadening was one half that of Br $\gamma$ 
profile. The CO emission component was normalized so that when it was added 
to the model spectrum, the depth of the minima in the CO features of SS Cyg 
were reproduced. Dashed vertical lines indicate the wavelengths of the
$^{\rm 12}$CO$_{(2,0)}$ (left) and $^{\rm 12}$CO$_{(3,1)}$ (right) bandheads.}
\label{sscygcocomp}
\end{figure}

If we assign the CO emission to the disk, then the broadening of any
spectral feature located there will be much greater than that of the slower 
rotational broadening of the secondary star (89 km s$^{\rm -1}$ in the
case of SS Cyg). For
example, if the accretion disk extends to 75\% of R$_{\rm L_{\rm 1}}$
(the Roche lobe radius of the white dwarf), using the stellar parameters
for SS Cyg in Bitner et al. (2007), the Keplerian velocity at this radius will 
be $\sim$ 400 km s$^{\rm -1}$. In Fig. \ref{sscygcocomp}, we add a
rotationally broadened CO spectrum to the best fitting synthetic spectrum
for SS Cyg using the value for the H I Br$\gamma$ line (FWHM  = 73 \AA), and 
a CO emission spectrum broadened by half that value. As could be expected,
the result is a dramatic change in the profiles of the CO features, with
an emission peak preceding each bandhead, a subtle redward shift for the 
location of the strongest absorption, and a dramatic narrowing of the CO
bandhead profiles. Obviously, the data for none of our objects resembles
this result. More complicated scenarios for the emission profile could
be constructed, but to truly diminish the CO features without distorting
the resulting spectrum {\it requires} the addition of CO emission with a 
broadening profile that is very similar to that of the secondary star. There is
no other location in a CV system besides the secondary star photosphere where 
such low velocities occur.

\section{Conclusions}
We have modified the spectral synthesis program MOOG and have demonstrated
that we are able to model moderate and lower resolution $K$-band spectra 
for stars hotter than T$_{\rm eff}$ $\geq$ 4000 K. We then used this program 
to investigate such spectra for three CVs: SS Cyg, RU Peg, and GK Per. It is 
clear that all three CVs in this sample have subsolar carbon abundances. That
synthetic spectra with identical parameters simultaneously model data with 
dramatically different dispersions for each of the objects, and realistic CO 
veiling scenarios fail to reproduce the observed CO features, proves that the 
carbon deficits are real.
It is important to note that these results represent the first direct
measurements of abundances in the photosphere of the mass donor stars in CVs. 
While our determinations of T$_\mathrm{eff}$, [Fe/H], and [C/Fe] for these
three objects are not overly precise, they have no parallel. There have not 
been many predictions for the behavior of the abundances in the atmospheres
of CV donor stars as a function of age/evolutionary state. Marks \& Sarna 
(1998) explored how the photospheric abundances of CNO elements in CV 
secondaries might
evolve with time, including the possibilities of pollution by the common
envelope phase, or through the accretion of classical novae ejecta. They
found that the latter two scenarios were inadequate in producing observable
changes, and concluded that only pre-contact evolution could significantly
alter the CNO abundances and isotopic ratios in CV secondary star photospheres.
Schenker et al. (2002) examined how CVs might evolve from supersoft binaries
and, if so, could show dramatic changes in CNO species. In light of our 
results, it would be useful to revisit this research with newer generations of 
binary star evolution codes, such as STAREVOL (Stancliffe \& Eldridge 2009). 
It is also obvious that moderate resolution spectroscopy of the brightest CVs 
covering the {\it entire} $K$-band are required to achieve more robust 
measures of [Fe/H], as well as allow for the proper examination of the CO 
bands. If the deficits of carbon arise from the CNO cycle, than the isotopic 
abundances of carbon will be altered. Moderate resolution $K$-band spectroscopy
covering the main $^{\rm 13}$CO features will allow for the confirmation of 
this inference.

\acknowledgements Both TEH and RTH were partially supported by a grant
from the NSF (AST-1209451). We acknowledge with thanks the variable star 
observations from the AAVSO International Database contributed by observers 
worldwide and used in this research. 

\begin{center}
{\bf References}
\end{center}
Allard, F., Homeier, D., \& Freytag, B. 2011, ASPC, 448, 91\\
Alves-Brito, A., Melendez, J., Asplund, M., Ramirez, I., et al. 2010, A\&A, 513, 35\\
Baron, E., Chen, B., \& Hauschildt, P. H. 2009, in AIP Conf. Ser., Vol. 1171, 
American Institute of Physics Conference Series, ed. I. Hubeny, J. M. Stone,
K. MacGregor, \& K. Werner, 148–160\\
Bitner, M. A., Robinson, E. L., \& Behr, B. B. 2007, ApJ, 662, 564\\
Bonifacio, P., Caffau, E., Ludwig, H.-G., \& Steffen, M. 2012, IAUS 282, 213\\
Cushing, M. C., Rayner, J. T., \& Vacca, W. D. 2005, ApJ, 623, 1115\\
Cushing, M., Vacca, W. D., \& Rayner, J. T. 2004, PASP, 116, 362\\
Dhillon, V. S., Littlefair, S. P., Marsh, T. R., Sarna, M. J., \& Boakes, E. H.
2002, A\&A, 393, 611\\
Dunford, A., Watson, C. A., \& Smith, R. C. 2012, MNRAS, 422, 3444\\
Duquennoy, A., Tokovinin, A. A., Leinert, Ch., Glindemann, et al. 1996, A\&A, 314,
846\\
Friend, M. T., Martin, J. S., Connon-Smith, R., \& Jones, D. H. P. 1990, 
MNRAS, 246, 654\\
Goorvitch, D. 1994, ApJS, 95, 535\\
Gray, D. F., The Observation and Analysis of Stellar Photospheres (Cambridge: 
Cambridge Univ. Press), 374\\
Gray, R. O., \& Corbally, C. J. 1994, AJ, 107, 742\\
Grevesse, N., Asplund, M., \& Sauval, A. J. 2007, Space Sci. Rev., 130, 105\\
Gustafsson, B., Bell, R. A., Eriksson, K., \& Nordlund, A. 1975, A\&A, 42, 407\\
Gustafsson, B., Edvardsson, B., Eriksson, K., J$\o$rgensen, U. G., Nordlund, A.,
\& Plez, B. 2008, A\&A, 486, 951\\
Hamilton, R. T. 2013, PhD Thesis, New Mexico State University\\
Hamilton, R. T., Harrison, T. E. Tappert, C., \& Howell, S. B. 2011, ApJ, 
728, 16\\
Harrison, T. E., Bornak, J., McArthur, B. E., \& Benedict, G. F. 2013, ApJ,
767, 7\\
Harrison, T. E., Howell, S. B., Szkody, P., \& Cordova, F. A. 2007a, AJ, 133,
162\\
Harrison, T. E., Campbell, R. K., Howell, S. B., Cordova, F. A., \& Schwope,
A. D. 2007b, ApJ, 656, 444\\
Harrison, T. E., Osborne, H. L., \& Howell, S. B. 2005, AJ, 129, 2400\\
Harrison, T. E., Osborne, H. L., \& Howell, S. B. 2004, AJ, 127, 3493\\
Hauschildt, P. H., \& Baron, E. 2010, A\&A, 509, 36\\
Hauschildt, P. H., Allard, F., Ferguson, J., Baron, E., \& Alexander, D. R. 
1999, ApJ, 525, 871\\
Howell, S. B., Harrison, T. E., \& Szkody, P. 2004, ApJ, 602, L49\\
Hummer, D. G., \& Storey, P. J. 1987, MNRAS, 224, 801\\
Kraft, R. R. 1964, ApJ, 139, 457\\
Kraft, R. P. 1962, ApJ, 135, 408\\
Ludwig, H. -G., Caffau, E., Steffen, M., Freytag, B., et al. 2009, MmSAI,
80, 711\\
Magic, Z., Collett, R., Hayek, W., \& Asplund, M. 2013, A\&A, 560, 8\\
Magic, Z., Collett, R., Asplund, M., Trampedach, R., et al. 2013, A\&A, 557, 
26\\
Marks, P. B., \& Sarna, M. J. 1998, MNRAS, 301, 699\\
Mart\'{i}nez-Arn\'{a}iz, R., Maldonado, J., Montes, D., Eiroa, C., \& 
Montesinos, B. 2010, A\&A, 520, 79\\
M\'{e}sz\'{a}ros, Sz., Allende Prieto, C., Edvardsson, B., Castelli, F., et al. 
2012, AJ, 144, 120\\
Morales-Reuda, L., Still, M. D., Roche, P., Wood, J. H., \& Lockley, J. J. 
2002, MNRAS, 329, 597\\
Politano, M. \& Weiler, K. P. 2007, ApJ, 665, 663\\
Reddy, R. R., \& Viswanath, R. 1990, Journal of Astrophysics and Astronomy, 
11, 67\\
Sbordone, L., Bonifacio, P., Castelli, F., \& Kurucz, R. L. 2004, Memorie 
della Societa Astronomica Italiana Supplementi, 5, 93\\
Scoville, N. Z., Krotkov, R., \& Wang, D. 1980, ApJ, 240, 929\\
Schenker, K., King, A. R., Kolb, U., Wynn, G. A., \& Zhang, Z. 2002, MNRAS, 337, 1105\\
Sion, E. M., Cheng, F., Godon, P., Urban, J. A., \& Szkody, P. 2004a, AJ, 128, 
1834\\
Sneden, C. A. 1973, PhD Thesis, University of Texas\\
Soubiran, C., Le Campion, J. -F., Cayrel de Strobel, G., \& Caillo, A. 2010,
A\&A, 515, 111\\
Sousa, S. G., Santos, N. C., Israelian, G., Mayor, M., et al. 2011, A\&A, 533,
141\\
Stancliffe, R. J., \& Eldridge, J. J. 2009, MNRAS, 396, 1699\\
Verbunt, F. \& Zwaan, C. 1981, A\&A, 100, 7\\
Wallace, K., \& Hinkle, K. 1996, ApJS, 107, 312\\
Wallace, K., Livingston, W., Hinkle, K., \& Bernath, P. 1996, ApJS, 106, 165\\
Warner, B. 1995, Cataclysmic Variable Stars (Cambridge: Cambridge Univ.
Press, 443\\
\clearpage
\begin{deluxetable}{lccccc}
\tabletypesize{\small}
\tablecolumns{6}
\tablewidth{0pt}
\centering
\tablecaption{Observation Log}
\tablehead{Object & Instrument & Date & Start Time & \#Exp. $\times$ 
Integration & Orbital Phase\\ &  & & (UT) & Time (s) & }
\startdata
SS Cyg & NIRSPEC & 2004 Aug. 28 & 12:31:51 & 4 $\times$ 300 & 0.20 \\
RU Peg & NIRSPEC &      "       & 13:17:12 & 4 $\times$ 300 & 0.87 \\ 
GK Per & NIRSPEC &      "       & 14:20:24 & 6 $\times$ 300 & 0.46 \\ 
$\gamma$ Psc& NIRSPEC &   " & 13:45:17 & 8 $\times$ 1 & \nodata\\
$\epsilon$ Eri& NIRSPEC &   " & 15:06:18 & 24 $\times$ 1 & \nodata\\
SS Cyg & SPEX    & 2004 Aug. 15 & 08:55:17 & 8 $\times$ 180 & 0.40\\
RU Peg & SPEX    &    "    & 10:25:06 & 8 $\times$ 240 & 0.85\\
GK Per & SPEX    &    "    & 13:53:54 & 6 $\times$ 240 & 0.94\\
\hline
\enddata
\end{deluxetable}

\begin{deluxetable}{lccccc}
\tabletypesize{\small}
\tablecolumns{6}
\tablewidth{0pt}
\centering
\tablecaption{Spectral Template Objects}
\tablehead{Object & Spectral Type  & Temperature (K) & log(g) & [Fe/H]& \# of Measures} 
\startdata
Sun           &G2V  &5777 & 4.44 & 0.0 & \nodata\\
Arcturus     &K0III &4313 $\pm$ 84 &1.65 $\pm$ 0.30& $-$0.54 $\pm$ 0.10& 34/31/31\\
$\epsilon$ Eri& K2V &5088 $\pm$ 72& 4.54 $\pm$ 0.16& $-$0.09 $\pm$ 0.09& 26/24/24\\
$\gamma$ Psc& G9III &4853 $\pm$ 45& 2.59 $\pm$ 0.26& $-$0.46 $\pm$ 0.12& 11/8/8\\
HD145675    & K0V  &5320 $\pm$ 114& 4.45 $\pm$ 0.07& $+$0.41 $\pm$ 0.12& 18/15/15\\
HD10476     & K1V  &5189 $\pm$ 53 & 4.46 $\pm$ 0.12& $-$0.06 $\pm$ 0.06& 16/10/10\\
HD3765      & K2V  &5023 $\pm$ 66 & 4.53 $\pm$ 0.16& $+$0.05 $\pm$ 0.10& 7/4/4\\
HD219134    & K3V  &4837 $\pm$ 138& 4.51 $\pm$ 0.13& $+$0.05 $\pm$ 0.10& 15/10/10\\
HD45977     & K4V  &4689 $\pm$ 174& 4.30 $\pm$ 0.39& $+$0.03 $\pm$ 0.18& 1/1/1\\
HD36003     & K5V  &4615 $\pm$ 29 & 4.35 $\pm$ 0.06& $-$0.14 $\pm$ 0.08& 3/2/2\\
HD201092    & K7V  &3964 $\pm$ 175& 4.50 $\pm$ 0.18& $-$0.26 $\pm$ 0.27& 10/5/5\\
\hline
\enddata
\end{deluxetable}

\begin{deluxetable}{lcccc}
\tablewidth{0.0pt}
\tablecolumns{4}
\tablecaption{Derived Secondary Star Parameters}
\tablehead{
  \colhead{System} &
  \colhead{$T_\mathrm{eff}$ (K)} &
  \colhead{log $g$} &
  \colhead{$\left[Fe/H\right]$\tablenotemark{\ }} &
  \colhead{$\left[C/Fe\right]$\tablenotemark{\ }} 
}
\startdata
SS Cyg & 4700 $\pm$ 250 & 4.50 & $-$0.25 $\pm $ 0.25 & $-$0.50 $\pm$ 0.25 \\
RU Peg & 4700 $\pm$ 250 & 4.50 & $+$0.00 $\pm$ 0.25 & $-$0.75 $\pm $ 0.25 \\
GK Per & 5100 $\pm$ 250 & 4.00 & $-$0.125 $\pm$ 0.25 & $-$1.00 $\pm$ 0.50 \\
\hline
\enddata
\end{deluxetable}


\clearpage

\clearpage

\clearpage

\clearpage

\clearpage

\clearpage

\clearpage

\end{document}